    \documentclass[twocolumn]{emulateapj}
    \usepackage{float,amsmath}
    \usepackage{graphicx}
    \usepackage{natbib}
    \usepackage{color}
    \newcommand{\volt}{{v}}
    \newcommand{\vis}{{V}}
    \newcommand{\sky}{{\rm sky}}
    \newcommand{\bmvolt}{{a}}
    \newcommand{\beam}{{A}}
    \newcommand{\thhat}{{\hat\theta}}
    \newcommand{\fngexp}{{e^{2\pi i\nu\vec{b}\cdot\thhat/c}}}
    \newcommand{\ifngexp}{{e^{-2\pi i\nu\vec{b}\cdot\thhat/c}}}
    \newcommand{\dfngexp}{{e^{2\pi i\nu \Delta \tau}}}

    \shorttitle{HERA dish reflectometry }
    \shortauthors{Patra et al.}
    
    \def\UCB{\altaffilmark{1}}
    
    \def\ASU{\altaffilmark{2}}

    \def\myemail{\altaffilmark{$\dagger$}}
    
\begin{document}
    
    \title{THE HYDROGEN EPOCH OF REIONIZATION ARRAY DISH III : MEASURING CHROMATICITY OF PROTOTYPE ELEMENT WITH REFLECTOMETRY. } 
    
    \author{
    Nipanjana Patra\UCB\myemail,
    Aaron~R.~Parsons\UCB,
    David~R.~DeBoer\UCB,
    Nithyanandan Thyagarajan\ASU,
    Aaron  Ewall-Wice\altaffilmark{7},
    Gilbert Hsyu\UCB, 
    T. K. Daisy Leung\UCB,
    Cherie~K.~Day\UCB,
    James E. Aguirre\altaffilmark{10}, Paul  Alexander\altaffilmark{5}, Zaki S. Ali\altaffilmark{1}, Adam P. Beardsley\altaffilmark{2}, 
   Judd D. Bowman\altaffilmark{2}, Richard F. Bradley\altaffilmark{8}, Chris L. Carilli\altaffilmark{9}, Carina  Cheng\altaffilmark{1}, Eloy  de~Lera~Acedo\altaffilmark{5}, Joshua S. Dillon\altaffilmark{1}, Gcobisa  Fadana\altaffilmark{13}, Nicolas  Fagnoni\altaffilmark{5}, Randall  Fritz\altaffilmark{13}, Steve R. Furlanetto\altaffilmark{4}, Brian  Glendenning\altaffilmark{9}, Bradley  Greig\altaffilmark{12}, Jasper  Grobbelaar\altaffilmark{13}, Bryna J. Hazelton\altaffilmark{14, 15}, 
   Daniel C. Jacobs\altaffilmark{2}, Austin  Julius\altaffilmark{13}, MacCalvin  Kariseb\altaffilmark{13}, Saul A. Kohn\altaffilmark{10}, Anna Lebedeva\UCB, Telalo  Lekalake\altaffilmark{13}, Adrian  Liu\altaffilmark{1, 16}, Anita  Loots\altaffilmark{13}, David  MacMahon\altaffilmark{1}, Lourence  Malan\altaffilmark{13}, Cresshim  Malgas\altaffilmark{13}, Matthys  Maree\altaffilmark{13}, Zachary  Martinot\altaffilmark{10}, Nathan  Mathison\altaffilmark{13}, Eunice  Matsetela\altaffilmark{13}, Andrei  Mesinger\altaffilmark{12}, Miguel F. Morales\altaffilmark{14}, Abraham R. Neben\altaffilmark{7},  Samantha  Pieterse\altaffilmark{13}, Jonathan C. Pober\altaffilmark{3}, Nima  Razavi-Ghods\altaffilmark{5}, Jon  Ringuette\altaffilmark{14}, James  Robnett\altaffilmark{9}, Kathryn  Rosie\altaffilmark{13}, Raddwine  Sell\altaffilmark{13}, Craig  Smith\altaffilmark{13}, Angelo  Syce\altaffilmark{13}, Max  Tegmark\altaffilmark{7}, Peter K.~G. Williams\altaffilmark{6}, Haoxuan  Zheng\altaffilmark{7}
    }
    
    \altaffiltext{1}{Department of Astronomy, University of California, Berkeley, CA}
    \altaffiltext{2}{School of Earth and Space Exploration, Arizona State University, Tempe, AZ}
    \altaffiltext{3}{Physics Department, Brown University, Providence, RI}
    \altaffiltext{4}{Department of Physics and Astronomy, University of California, Los Angeles, CA}
    \altaffiltext{5}{Cavendish Astrophysics, University of Cambridge, Cambridge, UK}
    \altaffiltext{6}{Harvard-Smithsonian Center for Astrophysics, Cambridge, MA}
    \altaffiltext{7}{Department of Physics, Massachusetts Institute of Technology, Cambridge, MA}
    \altaffiltext{8}{National Radio Astronomy Observatory, Charlottesville, VA}
    \altaffiltext{9}{National Radio Astronomy Observatory, Socorro, NM}
    \altaffiltext{10}{Department of Physics and Astronomy, University of Pennsylvania, Philadelphia, PA}
    \altaffiltext{11}{Department of Physics and Electronics, Rhodes University, PO Box 94, Grahamstown, 6140, South Africa}
    \altaffiltext{12}{Scuola Normale Superiore, Pisa, Italy}
    \altaffiltext{13}{SKA-SA, Cape Town, South Africa}
    \altaffiltext{14}{Department of Physics, University of Washington, Seattle, WA}
    \altaffiltext{15}{eScience Institute, University of Washington, Seattle, WA}
    \altaffiltext{16}{Hubble Fellow}
    
    \begin{abstract}
      The experimental efforts to detect the redshifted 21 cm signal from the Epoch of Reionization (EoR) are limited predominantly by the chromatic instrumental systematic effect. 
The delay spectrum methodology for 21 cm power spectrum measurements brought new attention to the critical impact of an antenna?s chromaticity on the viability of making this measurement. This methodology established a straightforward relationship between time-domain response of an instrument and the power spectrum modes accessible to a 21 cm EoR experiment. We examine the performance of a prototype of the Hydrogen Epoch of Reionization Array (HERA) array element that is currently observing in Karoo desert, South Africa. We present a mathematical framework to derive the beam integrated frequency response of a HERA prototype element in reception from the return loss measurements between 100-200 MHz and determined the extent of additional foreground contamination in the delay space. 
The measurement reveals excess spectral structures in comparison to the simulation studies of the HERA element. Combined with the HERA data analysis pipeline that incorporates inverse covariance weighting in optimal quadratic estimation of power spectrum, we find that in spite of its departure from the simulated response, HERA prototype element satisfies the necessary criteria posed by the foreground attenuation limits and potentially can measure the power spectrum at spatial modes as low as $k_{\parallel} > 0.1h$~Mpc$^{-1}$. The work highlights a straightforward method for directly measuring an instrument response and assessing its impact on 21 cm EoR power spectrum measurements for future experiments that will use reflector-type antenna.
    \end{abstract}
    
  \section{\textbf{Introduction}}
     Since it was first proposed in \cite{1997ApJ...475..429M, Shaver_et_al1999}, measurement of the redshifted 21~cm
    emission from the neutral hydrogen in the early Universe has gained attention as a
    powerful probe of both cosmology and astrophysics. The science case for
    21~cm cosmology, particularly during the Epoch of Reionization, is well
    established (see, e.g.,
    ~\cite{furlanetto_et_al2006, morales_wyithe2010, pritchard_loeb2012}).
    However, the technical path toward measuring this signal has been extremely challenging. The
    weakness of the hyperfine line keeps the 21~cm signal below the foreground throughout
    cosmological history and creates sensitivity and calibration
    challenges that are yet to be fully solved. With the system noise temperatures dominated
    by sky noise and foregrounds that are four to five orders of magnitude
    brighter than the signal even in the coldest patches of the sky, 
    sky-averaged 21~cm monopole experiments such as
    EDGES ~\citep{Bowman_et_al2010};
    SARAS ~\citep{Patra_et_al2013, Patra_et_al2015};
    SCIHI ~\citep{2015PhDT........65V};
    HYPERION ~\citep{presley_et_al2015};
    and experiments attempting to detect the 21~cm power spectrum such as
    the LOw Frequency ARray (LOFAR; ~\citealt{van_Haarlem_2013}),
    the Murchison Widefield Array (MWA; ),
    the Giant Metre-wave Radio Telescope (GMRT; ~\citealt{Paciga_et_al2011}),
    the Donald C. Backer Precision Array to Probe the Epoch of Reionization (PAPER; ~\citealt{parsons_et_al2010}),
    the Hydrogen Epoch of Reionization Array (HERA; ~\citep{deBoer_2016}),
    and the future Square Kilometre Array (SKA; \footnote{www.skatelescope.org}~\citealt{2013ExA....36..235M})
    must furnish their foreground suppression limits given their system performance at levels significantly
    beyond anything previously achieved in radio telescopes operating below 1 GHz.
    
    One of the most problematic effects these experiments face is the
    chromaticity that are instrument specific and have unique manifestation in the measured data.  The spectral dimension of    
    21~cm reionization experiments is of vital
    importance; for line emission, this coordinate translates to a line-of-sight distance
    that can be used to construct three-dimensional maps (and power spectra) of emission,
    as well as probe the evolution of 21~cm emission over cosmological timescales.
    The response of a radio telescope --- either a single dish
    or an interferometer --- versus spectral frequency, modulates spectrally smooth foregrounds,
    contaminating spectral modes that might
    otherwise be used to measure reionization ~\citep{ bowman_et_al2009, datta_et_al2010, trott_et_al2012, vedantham_et_al2012, liu_et_al2014a, 2012ApJ...752..137M}. Moreover, measurement sensitivity could be increased by increasing the collecting area of the element using reflector structures that  results in increased elementsize. However, the chromaticity of
  a reflector dish scales with its diameter, putting the
    needs of foreground suppression and signal sensitivity in direct tension with one
    another. 
    
    A major step forward for the field of 21~cm cosmology has been the development
    of a mathematical description of telescope chromaticity, how it varies with
    element separation (or telescope diameter), and what it implies for distinguishing
    the foreground emission from the cosmological 21~cm signal~\citep{Parsons_et_al_2012, vedantham_et_al2012, trott_et_al2012, 2015PhRvD..91b3002D, liu_et_al2014a,
   liu_et_al2014b, nithya_et_al2013,  Thyagarajan_et_al2015}.  The ``wedge'', as it is
    colloquially known, describes a linear relationship between the separation between elements
    in an interferometric baseline and the maximum line-of-sight Fourier mode\footnote{Assuming
    a flat sky and using appropriate
    cosmological scalars,
    the spectral axis, $\nu$, and angle on the sky, $\vec\theta$, translate to coordinates in
    a three-dimensional volume at cosmological distances.  In describing the spatial power spectrum of
    emission in this volume $P(\vec k)$, we use the three-dimensional wave vector 
    $\vec k\equiv(k_\parallel,\vec k_\perp)$, where $k_\parallel$ is aligned with the
    spectral axis, $\nu$, and $\vec k_\perp$ lies in the plane of the sky.}
    that may be occupied by smooth spectrum foreground emission.  At low-order $k_\parallel$ modes
    within the limits of the wedge,
    foreground contamination may be suppressed through a combination of calibration and model 
    subtraction.  However, calibration or modeling errors rapidly re-establish the characteristic
    wedge pattern~\citep{2016MNRAS.461.3135B}.  Outside of the wedge, foreground contamination drops rapidly 
    ~\citep{pober_et_al2013, 2014PhRvD..89b3002D,  Thyagarajan_et_al2015, Saul_et_al2016}, provided that the spectral responses
    of antenna elements and analog electronics are sufficiently smooth\footnote{Here, we distinguish
    the chromaticity of the elements in isolation from the chromaticity inherent to
    element separation in an interferometric baseline.}.  In the avoidance-based
    foreground strategy employed by PAPER ~\citep{parsons_et_al2014, Ali_et_al2015}, these modes
    may be targeted as the lowest risk path for constraining 21~cm reionization in the near term.
      
    The efficacy of foreground removal from the measured data
    critically limits the 21~cm power spectrum measurements. The delay
    transformation technique of foreground removal, introduced by
    ~\citep{Parsons_Backer_2009, Parsons_et_al_2012}, computes the Fourier
    transform of the visibility measured by an interferometer and produces a
    spectrum, referred to as the delay spectrum hereafter, which is a function of
    the geometric time delay corresponding to the physical length of the baseline
    between two antennas. For the visibilities measured over a wide field across a
    wide frequency bandwidth, the delay spectrum consists of the convolution of the instrument
    response with the foreground signal, and the 21cm power spectrum, hereafter
    referred to as the EoR signal.
    The technique exploits the smooth spectral characteristics of the foreground
    and for a 
     visibility measurement by an interferometer
    with a baseline $b$, confines the foreground contribution to the computed delay
    spectrum within the largest possible time delay, $\tau_{g} = b/c$, corresponding to
    the given baseline length. In reality, however, foregrounds show a response beyond the maximum geometric delay    
    due to broadband spectrum of the foreground sources and chromatic instrument response. For an ideal system, this 
    is a rapidly diminishing response beyond $\tau_{g}$ spanning the delay range corresponding to the inverse of the  
    measurement bandwidth. 
    Beyond this limit, for an ideal system performance, the delay spectrum would be
    dominated by any signal with spectral and spatial
    fluctuations over small scales, such as the EoR signal.  The interaction between the sky signal and instrument response can
    alter the relative contribution of the foreground, and the
    EoR signal at a given delay and influence the detectability of the EoR
    signal. Such interactions may cause the foreground to spill
    over into higher delays and push the upper limit of the foreground and
    systematics contaminated delay modes to much higher delays affecting the EoR window. 
    
\begin{figure*}[ht]
     \includegraphics[ width = 0.9\linewidth]{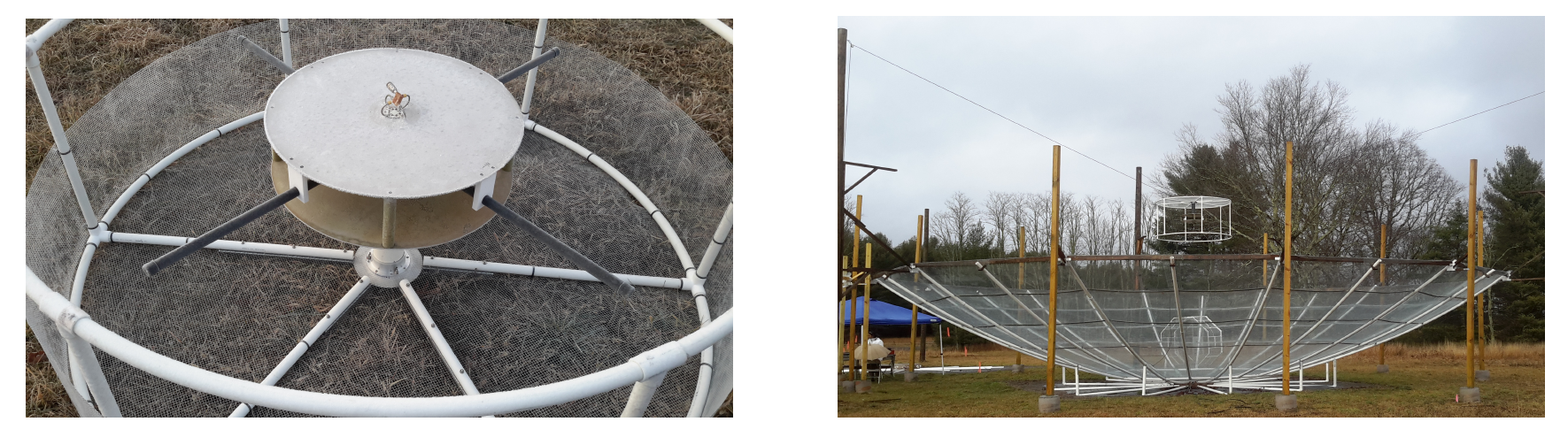}
   \caption{Left: The HERA feed consists of a pair of crossed-dipoles over 1.72 m diameter backplane made of wire mesh. The backplane is surrounded by 0.36 m wide wire mesh around the edge resulting in encasing the crossed-dipoles in a cylindrical cage.. Right: A prototype HERA element at the Green Bank NRAO site consisting of a crossed dipole feed and a parabolic reflector dish.}
    \end{figure*}
    
    In this paper, we use reflectometry measurements to
    investigate the delay-domain performance of 
    a prototype 14-m parabolic reflector dish and a crossed dipole feed (figure \ref{fig:heradish}) that are proposed to be used in HERA, operating 
    from 100 to 200~MHz. For clarity, the return loss of the feed when measured off of the dish will be referred to as ``feed return loss" hereafter. The dish-feed assembly, when the feed is placed at the focal point of the dish is referred to as ``HERA element" hereafter. 
     
    This paper is one in a series of five papers that studies the time and frequency domain simulation of  HERA element~\citep{Ewall-Wice_et_al2016}, ~\citep{ddboer_et_al2016}, beam pattern measurements~\citep{Neben_et_al2016}, and foreground limits ~\citep{Thyagarajan_et_al2016}. We compare 
    reflectometry measurements to the simulated specification for the spectral performance of a HERA element ~\citep{ddboer_et_al2016} and to the limits of EoR to foreground power ratio established by ~\citep{Thyagarajan_et_al2016}. 
    In this paper, Section 2 briefly describes the delay spectrum for the ideal and
    non-ideal performance of a two element interferometer. Section 3 describes the
    reflectometry measurements and establishes the connection between visibility
    measurements and reflectometry. Reflectometry results are described in Section
    5. Section 6 evaluates the performance of the HERA element for detection of the
    21~cm power spectrum. 

  \section{\textbf{Visibility measurements by a two element interferometer: the delay spectrum}}

   For a two element interferometer with a baseline $\vec b$ and antenna voltage gain pattern $\bmvolt_{1}$, $\bmvolt_{2}$, the voltage at the output of each antenna element can be written as, 
     \begin{equation}
    \volt_{1}(\thhat, \nu) = \bmvolt_{1}(\thhat,\nu)~\volt_\sky(\thhat,\nu)\nonumber\\
    \end{equation}
    \begin{equation}
    \volt_{2} (\thhat, \nu)= \bmvolt_{2}(\thhat,\nu)~\volt_\sky(\thhat,\nu)~\fngexp\nonumber\\
    \end{equation}
    
    where, $\volt_\sky(\vec \theta, \nu)$ is the sky voltage in the direction $\thhat$ at a frequency $\nu$. 
    The measured visibility is, 
    \begin{eqnarray}
    \vis(\vec b, \nu) & = & \int  \volt_{1}(\thhat,\nu)~  \volt_{2}^{*} (\thhat, \nu)~ \ifngexp d\Omega \nonumber\\
    			    & = &\int \beam(\thhat, \nu)~ I_\sky(\thhat,\nu)~ \ifngexp d\Omega
    \label{eq1}
    \end{eqnarray}
    
    where  $\beam(\thhat,\nu)=\bmvolt_{1}(\thhat,\nu)~\bmvolt_{2}^{*}(\thhat,\nu)$ is the cross power pattern and  $I_\sky(\thhat,\nu)=\volt_\sky(\thhat,\nu)~\volt_\sky^{*}(\thhat,\nu)$ is the sky intensity. Hence, 
    The Fourier transform of the visibility along the frequency axis i.e, the delay spectrum is,
    \begin{eqnarray}
     \tilde V(\vec b, \tau)  & = & \beam(\vec b,\tau)*I_\sky(\vec b,\tau)	   \nonumber\\
   	 & = & \int_{-\infty}^{\infty} \beam(\tau-\Delta \tau) I_\sky(\tau)d\Delta \tau 
\label{eq3}
\end{eqnarray}

for a given baseline $\vec b$ ~\citep{parsons_et_al2012a}.  
     The delay-transformed visibility is related to the power spectrum of redshifted
    21 cm emission by the relation, 
    \begin{equation}
 P_{21}(k_\perp,k_\parallel) \approx | \tilde V(\vec b, \tau)|^{2} \left(\frac{\lambda^2}{2k_\textrm{B}}\right)^2 {X^{2}Y \over \Omega \Delta B}
    \label{eq4}
    \end{equation}
    where
    \begin{equation}
      k_\perp = \frac{2\pi f}{D}\Bigg({b\over c}\Bigg), \text{ }
      k_\parallel = \frac{2\pi\tau\,f_{21}H(z)}{c(1+z)^2}, 
     \label{eq5}
    \end{equation}
   
    \begin{itemize}
    \item
     $f_{21}, f, z, \Delta B $: rest frame, observation frequency of the 21~cm radiation and redshift and bandwidth of observation.
     \item
     $k_\textrm{B}$: Boltzmann constant.
     \item
     $D\equiv D(z)$ comoving distance along the line of sight
     \item
    $H(z)= H_{0} [\Omega_\textrm{M}(1+z)^3+\Omega_\textrm{R}(1+z)^2+\Omega_\Lambda]^{1/2}$\\
     is the Hubble constant as a function of redshift, $H_{0}= 100 h$~km s$^{-1}$ Mpc$^{-1}$, $\Omega_\textrm{M}=0.27, \Omega_\textrm{R}= 0.73, \Omega_\Lambda=1-\Omega_\textrm{M}-\Omega_\textrm{R}$ are the matter, radiation and dark energy density parameters respectively.
    \item
    $X, Y$ are cosmological scalars relateing the angular dependence with spectral frequency to corresponding comoving distance.
    \end{itemize}

    \section{\textbf{Effects of multiple reflections on visibility and delay spectrum}}
    
    HERA consists of parabolic reflector antennas which
    provide increased collecting area per array element compared to its predecessor
    experiment PAPER. Plane waves incident on a prime focus parabolic dish are focussed at the
    feed which is at the focal plane of the dish.  The mismatch between the
    impedance of the free space with the feed and transmission line results in a partial
    coupling of the sky signal while the rest is reflected off of the
    feed.  This signal illuminates the dish and most of it is reflected
    back into the space.  However, a part of it reflects back and forth several
    times between the feed and the vertex of the dish which is shadowed by the
    feed.  Such reflections generate multiple copies of the incident sky signal of
    reduced strength at various delays and phases producing additional
    correlations in the visibilities data.  Identical
    response can result from reflections internal to the system,
    for example between antenna output and backend receiver input causing similar signal contamination in delay space.  We measure the system response to 
    estimate these reflections and the  resulting effects in power spectrum measurements. 
    
     We assume that the reflection coefficient of the parabolic dish is proportional to its area of illumination i.e, upon complete illumination of the dish $100\%$ of the incident radiation is reflection off of the dish and reflection coefficient of the dish is unity. The sky voltage incident on the dish illuminates the part of the dish that is not shadowed by the feed. The corresponding dish reflection coefficient is denoted as $(1-\Gamma_{d})$ which is complex. The dish reflection coefficient of the part of the dish shadowed by the feed is denoted by$\Gamma_{d}$.
     The voltage reflection coefficient of the feed is $\Gamma_f$ and corresponding transmission coefficient is $(1+\Gamma_f)$ (see, {\em e.g.} Pozar).  Both $\Gamma_{d}$ and $\Gamma_{f}$ are complex numbers that  
    are functions of frequencies. For a given incidence of the sky voltage
    $\volt_{sky}$,  $\Gamma_{d}(1+\Gamma_{f})$ fraction of it is
    coupled into the cable leading to the receiver backend from the feed whereas $\Gamma_f$ fraction of it is reflected off. This subsequently undergoes $\Gamma_d$ reflection off the dish and $(1+\Gamma_{f})$ of it reenters the feed with a
    roundtrip time delay $\Delta \tau=2F/c$ where F is the focal length of the dish. Hence, if $v_{sky}$ is reflected $n$
    times in between the feed and the dish, the net voltage entering the feed after
    the $n^{th}$ reflection can be written as:
    \begin{eqnarray}
    \volt_{rec} & = &  (1-\Gamma_d) (1+\Gamma_{f}) \volt_{sky}[1+ \Gamma_{f}\Gamma_{d} \dfngexp \nonumber \\
    	&& + (\Gamma_{f}\Gamma_{d})^2  (\dfngexp)^{2}+ \nonumber \\
    &&  ....+ (\Gamma_{f}\Gamma_{d})^{n} (\dfngexp)^{n}]
    \label{eq6}
    \end{eqnarray}
    
    \begin{figure}[ht]
    \centering
    \includegraphics[width=\linewidth]{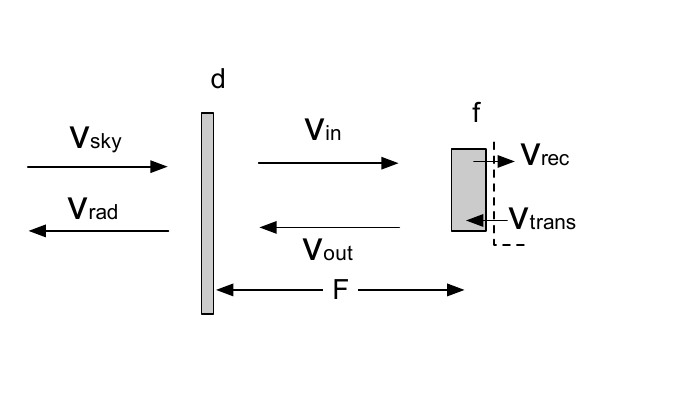}
    \caption{Schematic diagram of signal propagation through the dish and the feed of during both transmission and reception. $V_{sky}$, $V_{in}$ and $V_{rec}$ represents the sky voltage, total reflected voltage from the dish that is incident on the feed and the net received voltage at the receiver input. In transmission mode, $V_{trans}$, $V_{out}$ and $V_{rad}$ are the voltage transmitted by the network analyser, the voltage output of the feed and the the net radiated voltage after reflection on the dish. $\Gamma_{d}$,  $\Gamma_{f}$ are the voltage return loss of the dish and the feed. }
    \label{fig:sys}
    \end{figure}
    
    \noindent
    or,
     \begin{eqnarray}
    {\volt_{rec}\over \volt_{sky} } & = &  (1- \Gamma_d)(1+\Gamma_{f}) \displaystyle\sum\limits_{n=0}^{N} [\Gamma_{f}\Gamma_{d}\dfngexp]^{n}\nonumber\\
    \label{eq7}
    \end{eqnarray}
    Note that $n=0$ is when the initial wave enters the cable at the feed (where we assume we have set our reference plane). 

    An accurate measurement of this quantity requires receiving $v_{sky}$ from a
    well calibrated, wideband source in the sky in the far field of the HERA
    antenna element with significant isotropic emission. While this condition is
    hard to achieve in practice, reciprocity of the antenna performance
    in the transmission and reception mode implies the right hand side of equation
    \ref{eq7} could be measured using the return loss measurement technique with a
    vector network analyzer (VNA). \\
    \section{\textbf{Reflectometry measurements}} 
    Reflectometry measurements determine the multiple reflection of any signal within 
    a system in either time ~\citep{2015arXiv150205862P} or frequency domain. 
    We carried out frequency domain reflectometry on a prototype HERA element in
    Green Bank, WV (figure \ref{fig:heradish}) in order to measure the
    response of the feed and dish assembly. The prototype HERA
    element consists of a $14.5~m$ diameter parabolic reflector and a
    crossed-dipole pair as a feed. The cross dipole antenna pair is identical to
    the feed of the PAPER antenna and suspended at the focal plane of the dish
    with the support of three vertical poles. HERA elements will be closely spaced with
    centre to centre distance between the two adjacent dishes slightly larger than the dish diameter.
    The crossed
    dipole feed is encased in a cylindrical cage with the back plane to reduce the coupling between the adjacent dishes. The
    feed is raised and lowered by a pulley system mounted on the three poles.
    The focal height of the dish is $\approx 5$~m.  The detail geometry and electromagnetic
    design of of the feed is presented in \cite{ddboer_et_al2016}. 
 
    During our measurements, each HERA element is connected to an active balun similar
    to the ones used in the PAPER antenna which provides greater than 10dB return
    loss of power at the feed output.
  The active balun is replaced by a passive  one with 4:1 impedance ratio to match the impedance of the 50~$\Omega$ transmission line. These measurements, therefore, provide a conservative estimate for $\Gamma_{f}$ and $S_{11}$.   
  \subsection{\textbf{Measurement equations}}
   The measurement presented here is done using a vector network analyzer (VNA) that is connected to the antenna using a 15~m long LMR 400 cable. This length correspond to a 120~ns roundtrip delay. 
     A VNA is connected to the HERA element via a $\approx 15$~m cable that transmits
    a broadband noise voltage, $\volt_{trans}$, and a factor $(1+\Gamma_f)$ of this
    voltage is radiated by the feed while the fraction $\Gamma_{f}$ returns to the
    VNA. The transmitted signal illuminates the dish and most of it is
    radiated into the free space, a fraction $\Gamma_d$ of this signal is reflected
    back towards the feed, and $(1+\Gamma_f)$ fraction of it is received by the VNA.  The received
    voltage, after $n$ reflections between the dish and the feed (where again $n=0$
    is the first reflected signal at the reference plane), is therefore:
    \begin{eqnarray}
    \volt_{rec} & = &  \Gamma_f \volt_{trans} \nonumber \\
             && + \volt_{trans} (1+\Gamma_f)^2 \Gamma_{d} \dfngexp \nonumber \\
             && + \volt_{trans} (1+\Gamma_f)^2 \Gamma_{d} \dfngexp \Gamma_d\Gamma_f\dfngexp \nonumber \\
             && + \volt_{trans} (1+\Gamma_f)^2 \Gamma_{d} \dfngexp (\Gamma_d\Gamma_f\dfngexp)^2 \nonumber \\
    &&  ....+ \volt_{trans} (1+\Gamma_f)^2 \Gamma_{d} \dfngexp (\Gamma_d\Gamma_f\dfngexp)^n \nonumber \\
    \label{eq8}
    \end{eqnarray}
    or, 
    \begin{equation}
    {\volt_{rec}\over \volt_{trans} } = \Gamma_f + \frac{(1+\Gamma_f)^2}{\Gamma_{f}} \displaystyle\sum\limits_{n=1}^{N} [\Gamma_{f}\Gamma_{d}\dfngexp]^{n}
    \label{eq9}
    \end{equation}
    The VNA measures the quantity $\volt_{rec}/\volt_{trans}=s_{11}$ which is the voltage reflection coefficient of the HERA element.
    The fundamental difference between the equations \ref{eq7} represents the system response in reception mode while our measurements are done in transmission mode (equations \ref{eq7}). The $n=0$ term in equations \ref{eq7} represents the transmission of the beam integrated sky signal from the antenna to the transmission line while the same in equations \ref{eq9} represents the reflection at the antenna terminal and the transmission line. The
    zero delay response is identical whether or not the feed is suspended on the
    dish. Therefore, we measure the feed return loss
    alone while the feed is kept on the ground facing the sky. This measurements
    include the reflections from the surrounding cage structure but exclude the
    dish response. 
     Using the
    return loss measurement of the feed $(\Gamma_{f})$, the return loss measurement
    $s_{11}$ of the dish and feed assembly is corrected via equation \ref{eq11}. 
    Writing $\volt_{rec}/\volt_{trans}=s_{11}$, equation \ref{eq9} can be written as,
    \begin{equation}
    s_{11} +\frac{(1+\Gamma_f)^2}{\Gamma_f}-\Gamma_f = \frac{(1+\Gamma_f)^2}{\Gamma_{f}} \displaystyle\sum\limits_{n=0}^{N} [\Gamma_{f}\Gamma_{d}\dfngexp]^{n}
    \label{eq10}
    \end{equation}
    Finally, 
    \begin{eqnarray}
    {\volt_{rec}\over \volt_{sky} } & = & (1-\Gamma_d) \left[(1+\Gamma_f) + \frac{\Gamma_f}{(1+\Gamma_f)}\left(s_{11} - \Gamma_f\right)\right] \nonumber\\
    \label{eq11}
    \end{eqnarray}
    %
    \indent From this, for a two element interferometer with identical antenna elements, the ratio of the received sky intensity to true sky intensity will be, 
    \begin{eqnarray}
    {I_{rec}\over I_{sky} } & = & \Bigg|{\volt_{rec}\over \volt_{sky} }\Bigg|^2 =  |1-\Gamma_d|^{2}
    \times \nonumber\\
                 && [ |1+\Gamma_f|^2 +  2\operatorname{Re}\left(\frac{\Gamma_f}{(1+\Gamma_f)}(s_{11} - \Gamma_f)\right)  \nonumber\\ 
                 &&  + \frac{|\Gamma_f|^2}{|1+\Gamma_{f}|^2}|s_{11} - \Gamma_f|^2]    \nonumber\\
    \label{eq12}             
    \end{eqnarray}
    
    In terms of visibility, the same could be written as,  
    \begin{eqnarray}
    \vis^{mul}(\vec b,\nu) & = & \int |1-\Gamma_d|^{2}\times \nonumber\\
                 && \left[|1+\Gamma_f|^2 +  2\operatorname{Re}\left(\frac{\Gamma_f}{(1+\Gamma_f)}(s_{11} - \Gamma_f)\right)\right] + \nonumber\\ 
                 &&  \left[ \frac{|\Gamma_f|^2}{|1+\Gamma_{f}|^2}|s_{11} - \Gamma_f|^2  \right]  I_{sky} \ifngexp d\Omega
    \label{eq13}
    \end{eqnarray}
    Comparing equations \ref{eq13} and \ref{eq1}, the cross power generated by
    $v_{1}$ and $v_{2}$ has a spurious visibility response due to mutual
    correlation between multiply reflected voltages represented by the second and
    the third term of the right hand side of equation \ref{eq13}. In the ideal
    case, $\Gamma_{f}=0$, in which case equation \ref{eq13} results in equation
    \ref{eq1}. The Fourier transform of this visibility spectrum along the
    frequency axis results in the delay spectrum.  In the delay domain,
    visibilities contributed by any two voltage components from two antennas
    with no mutual delay will be located at the delay $\tau = 0$ whereas any two
    voltage components from two antennas having a mutual delay of $n\Delta \tau$
    will be centred at $n\Delta \tau$. These will result in leakage of the foreground 
    to delays where EoR power spectrum could be measured.
    
     The delay transform of the measured return loss is sensitive to the bandwidth of measurements. Wideband 
    measurements over finite bandwidth is analogous
     to windowing the frequency domain data by a square
    window function that results in multiple side lobes at higher delays. We apply the Blackman-Harris window function of \citep{nithya_et_al2013, Thyagarajan_et_al2016} to estimate the delay transform.

    \subsection{\textbf{Results}}
\begin{figure*}[ht]
    \begin{minipage}[b]{0.5\linewidth}
    \centering
    \includegraphics[angle=0, width=\linewidth]{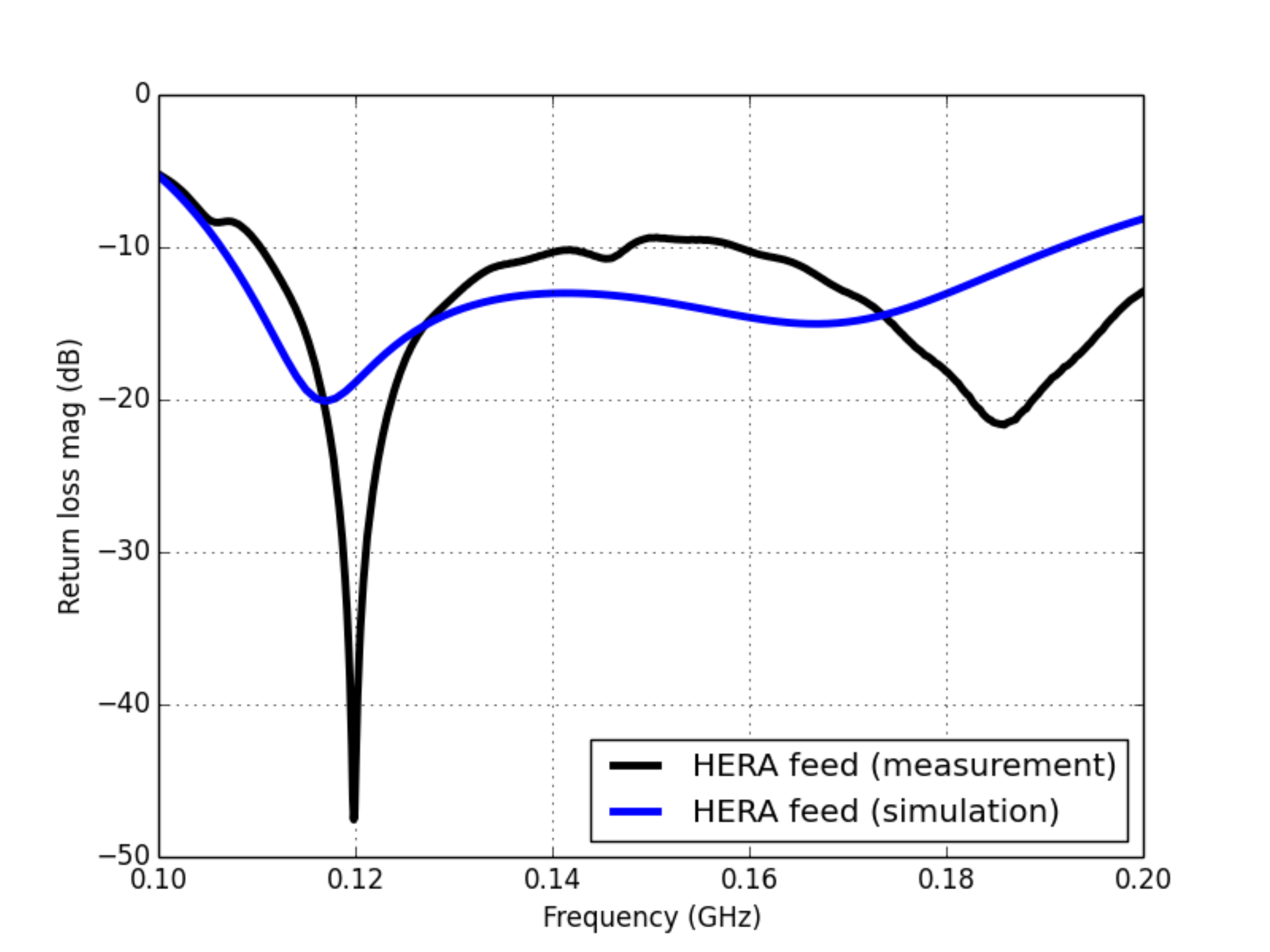}
    \end{minipage}
    \hspace{0.1cm}
    \begin{minipage}[b]{0.5\linewidth}
    \centering
    \includegraphics[angle=0, width=\linewidth]{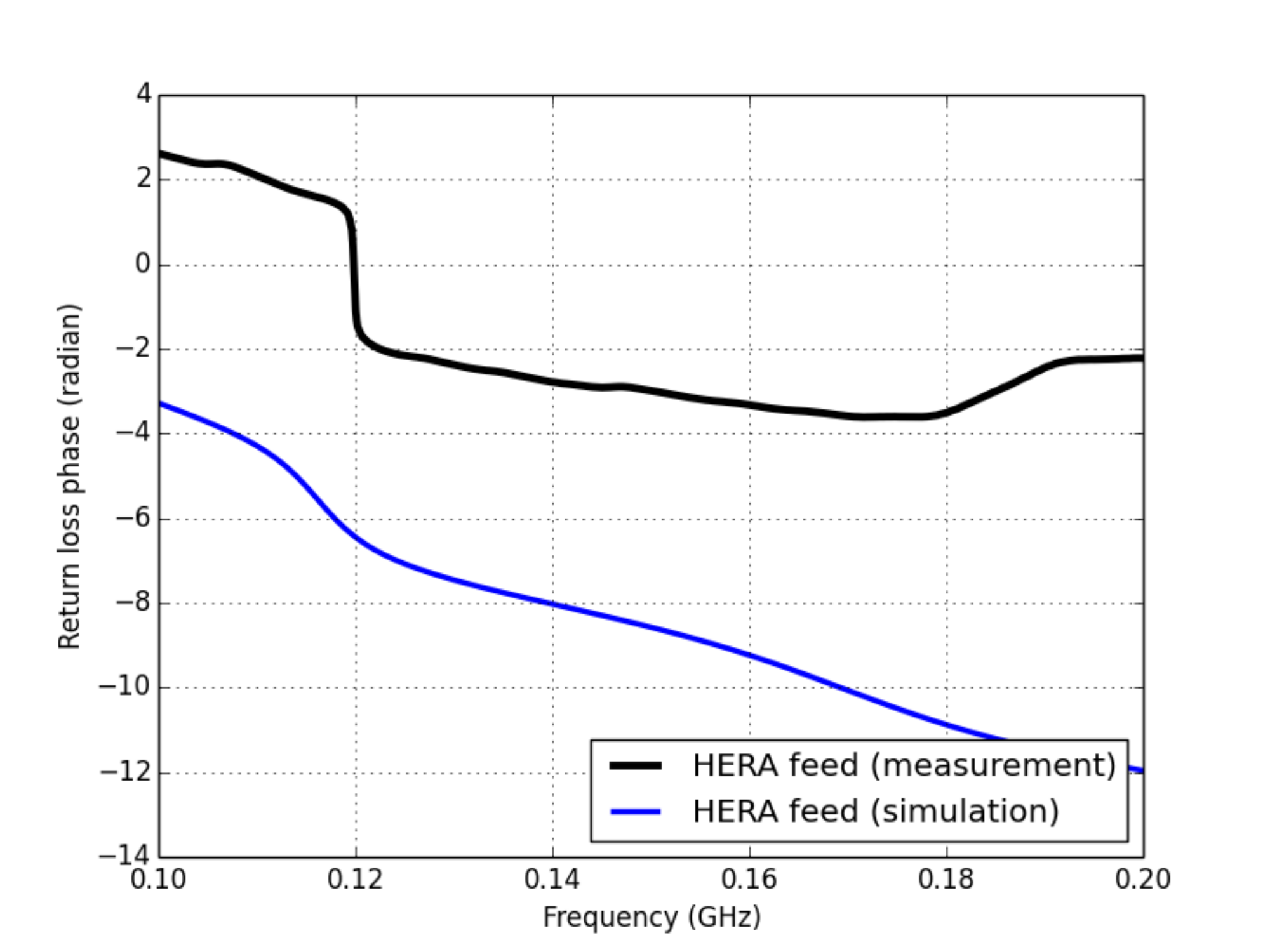}
    \end{minipage}
    \vspace{0.1cm}  
    \begin{minipage}[b]{0.5\linewidth}
    \centering
    \includegraphics[angle=0, width=\linewidth]{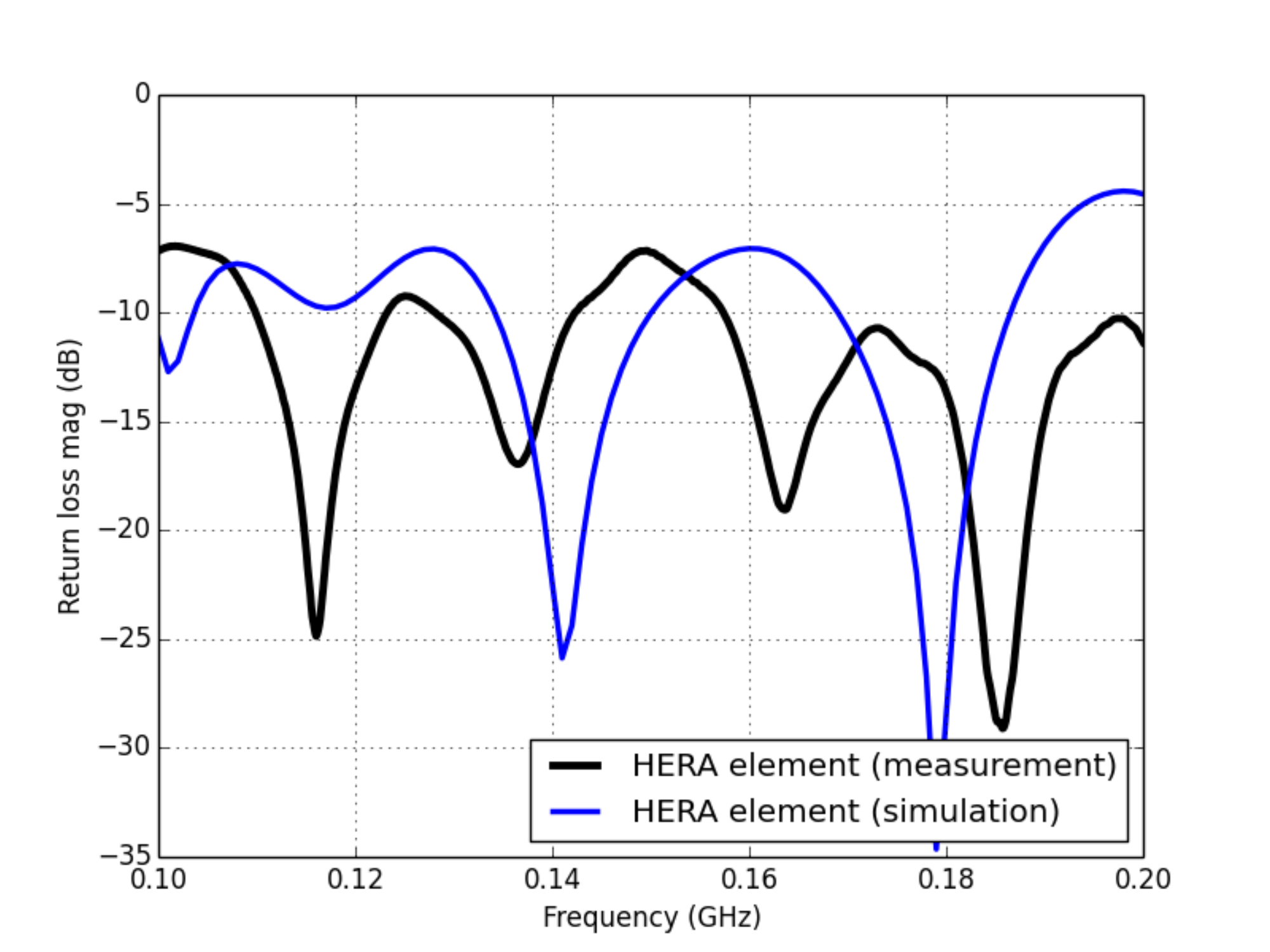}
    \end{minipage}
    \hspace{0.1cm}
    \begin{minipage}[b]{0.5\linewidth}
    \centering
    \includegraphics[angle=0, width=\linewidth]{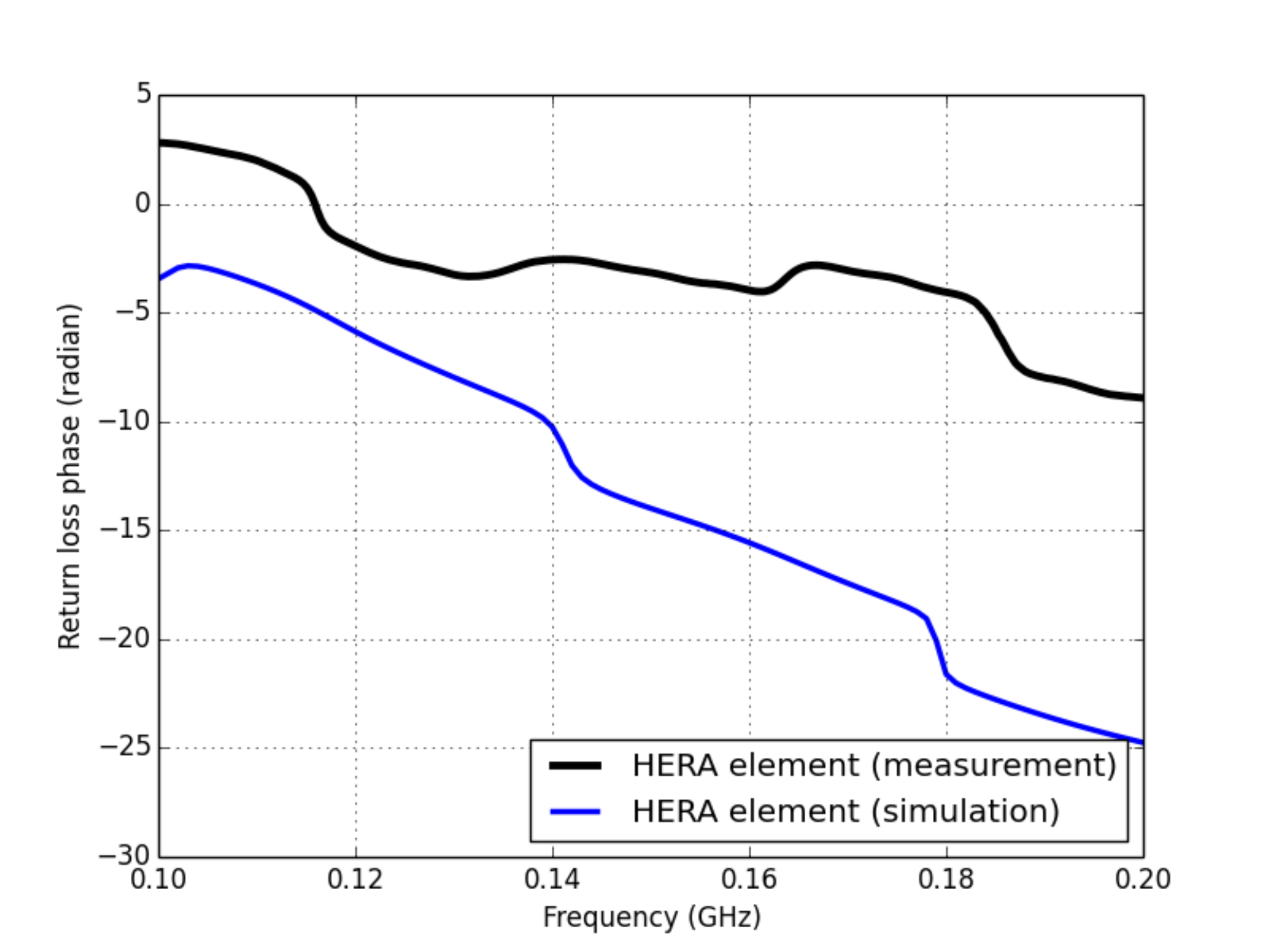}
    \end{minipage}
    \caption{Upper panel: Magnitude and phase of the return loss of the feed as simulated and measured. Both measurement and simulation shows similar level of return loss across the band with marginally better return loss at the high frequency end of the band. Lower panel: Magnitude and phase of return loss when the feed is suspended at the focal point of the dish which 5~m above the dish vertex.}   
    \label{RL_mag_dish}
    \end{figure*}

\begin{figure}[ht]
    \centering
    \includegraphics[angle=0, width=\linewidth]{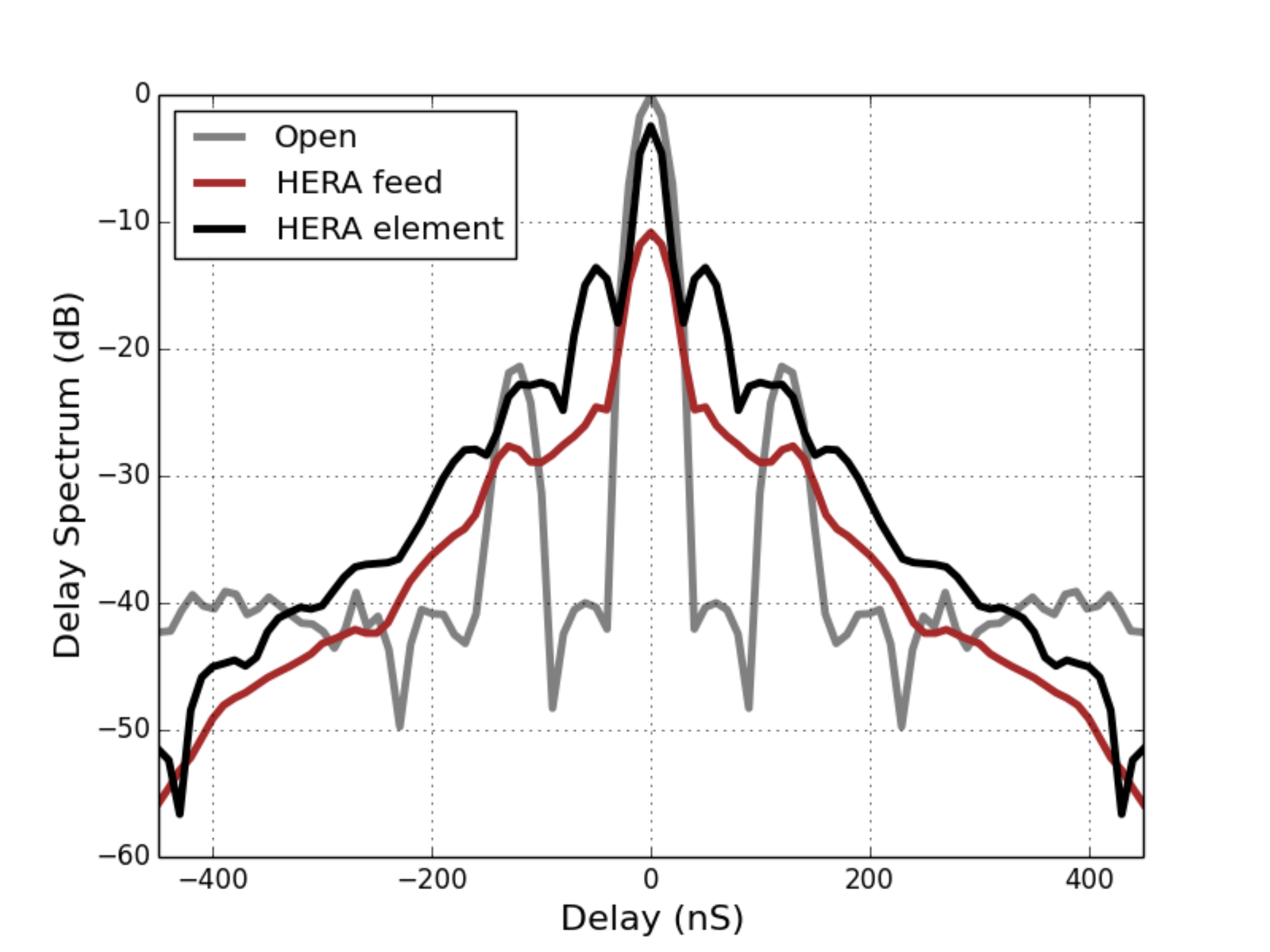}
    \caption{Delay spectrum of the HERA feed (brown) estimated by taking the Fourier transform of the measured return loss of power at the feed output (upper panel in figure \ref{RL_mag_dish}). The feed, when suspended on the dish results in a more complex return loss as shown in the lower panel of figure \ref{RL_mag_dish} and corresponding delay response is shown here in black. A balun with impedance transformation ratio $4:1$ and input return loss of -22~dB across the band is used for converting the antenna balanced output to unbalanced output voltage. Balun response is embedded in the measurements. Delay spectrum of the open load (gray) shows the reflections at the VNA input. }
    \label{ds_feed_on_dish_trans}       
    \end{figure}

\begin{figure}
    \centering
    \includegraphics[width=\linewidth]{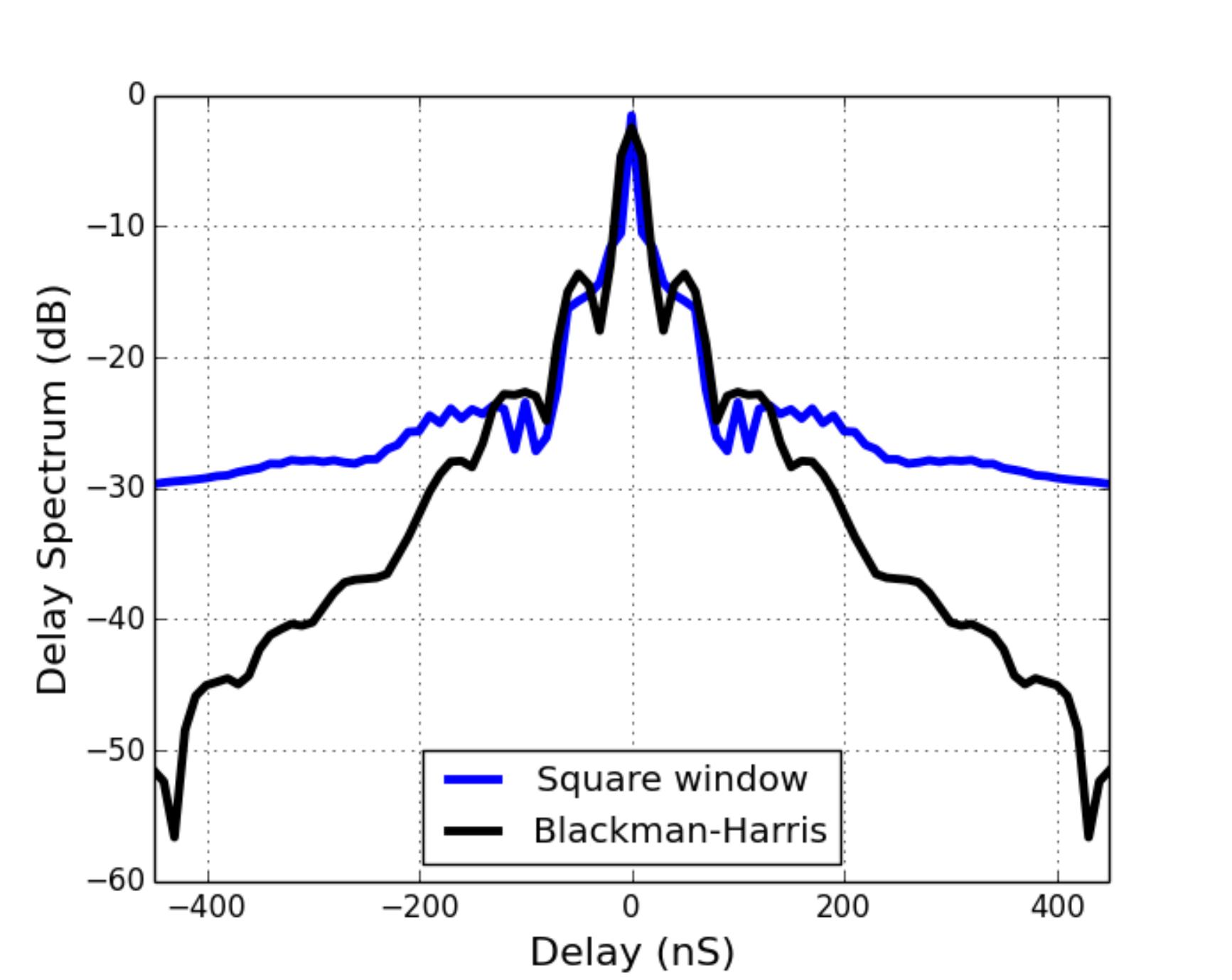}
    \caption{Effect of finite bandwidth on estimation of delay spectrum: Blue line shows the delay spectrum of the HERA element computed from the measured data which is band limited between 100 to 200~MHz. Black line shows the delay spectrum estimated from the same data set after multiplying the data by a Blackman Harris window. The delay spectrum of the windowed data set shows significant reduction in the instrument response at higher delays.}
    \label{fig:window}
    \end{figure}

    Return loss measurements of the feed as well as HERA element is shown in figure \ref{RL_mag_dish}. Corresponding        
    delay spectra are shown in figure \ref{ds_feed_on_dish_trans}. The delay spectra of the feed and the HERA element closely follow each other up to delays $< 50$~ns. Figure \ref{fig:window} shows the delay spectra estimated using a square window as well as Blackman-Harris window. In this case, the delay spectra follow each other up to $50$~ns and start deviating after that. This is because at small delays the spectrum is dominated by the small scale reflections associated with the feed cage. Above $50$~ns, the windowing effect begins to manifest. The relevant length scales inside the system from where the reflections are expected and the corresponding delays are summarized in the table \ref{delay_table}.  The delay spectrum of the feed exclude the dish reflections. However, the VNA reflections at the VNA input are present in both the cases. 
    We measure the VNA input reflections by connecting an
    open load at the feed input of the cable that results in $100\%$ signal reflection. The combined effect of the VNA input return loss
    along with the cable resistive loss is shown in figure
   (figure \ref{ds_feed_on_dish_trans}). Since our measurement plane is at the open end of the cable,
    first reflection appears at zero delay while the second one appears around 120~ns. 
    which is $\approx -22$~dB. Third and 
    consecutive reflections are buried in the noise floor of this measurement. Therefore, if the return loss at the backend  input is  $> -22$~dB, the delay spectrum contamination  due to the reflection at the receiver input will be negligible. Moreover, if the cable length is sufficiently increased
    without the loss of signal, these reflections could be made to occur at delays
    which are not of interest for EoR measurements.

\begin{table}
\caption{System delay scale}
\centering
\begin{tabular}{ccc}
\hline\noalign{\smallskip}
Roundtrip Delay (ns) & Length scale (m) & Source  \\
\noalign{\smallskip}\hline\noalign{\smallskip}
30 & 5 & Dish apex \\
2.86 & 1.72 & Feed Cavity \\
120 & 15 & VNA Input \\
\noalign{\smallskip}\hline
\end{tabular}
\label{delay_table}
\end{table}

    \subsection{\textbf{Comparison with simulations}}
  \begin{figure}
  \centering
  \includegraphics[width=\linewidth]{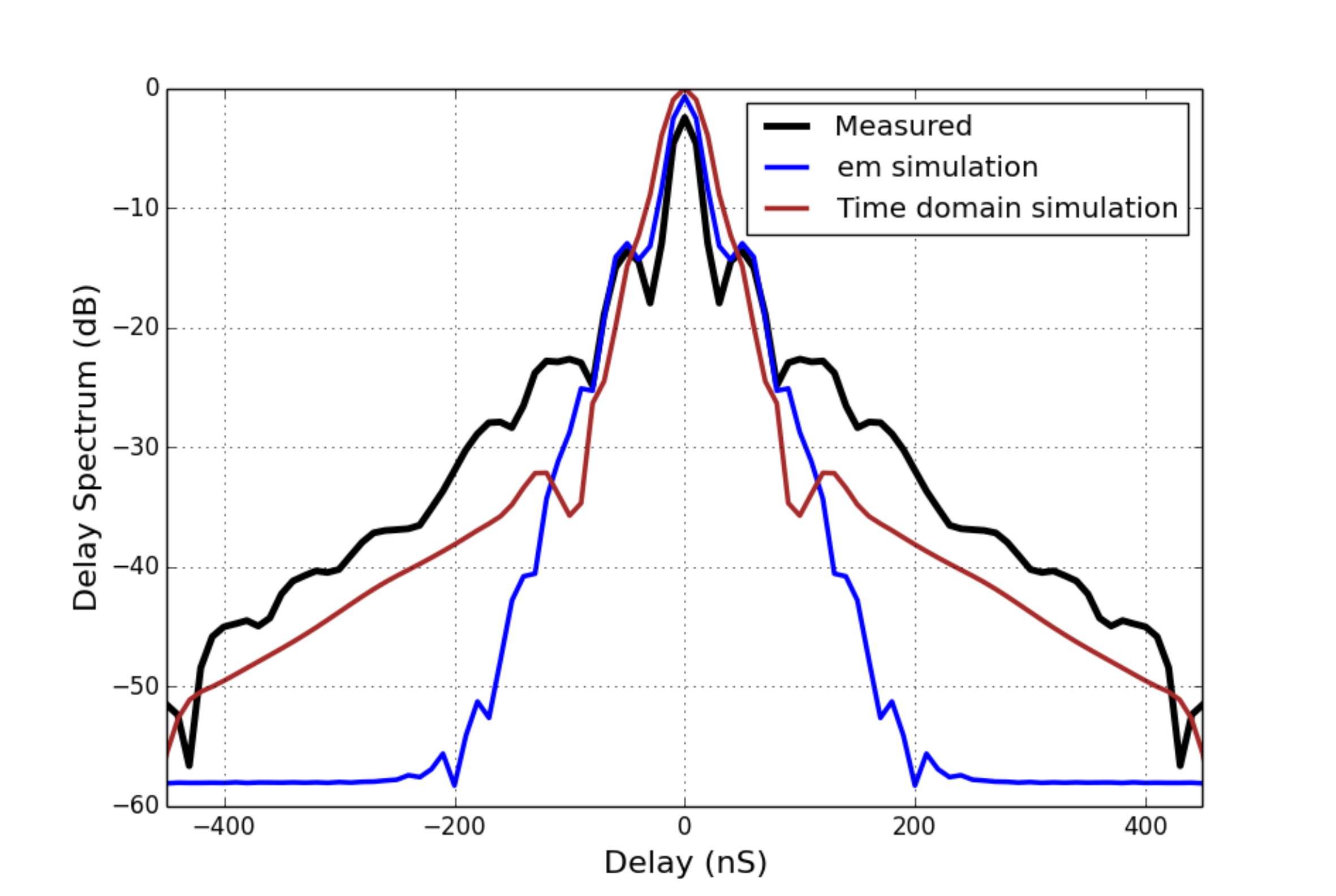}
  \caption{Delay spectrum of HERA element estimated from the reflectometry measurements by a vector network analyser compared to the delay spectrum estimated from the EM simulation using HFSS and time domain simulation using CST .}
   \label{fig:sim_em}
   \end{figure}
   
 \subsubsection{HFSS Electromagnetic simulation}  
 Performance of the HERA feed and the HERA element has been simulated by
    using the EM solver "High Frequency Structural Simulator (HFSS)" using the finite
    element method. These simulations provide a ballpark estimate of the system response under various ideal conditions. The measurement is in agreement with the simulation in some aspects and differ in some.   
    The simulated return loss of the feed by HFSS, both
    magnitude and phase are shown in the upper panel of figure
    \ref{RL_mag_dish} along with the measurements.  Bottom panel of figure \ref{RL_mag_dish} shows the same when
    feed is suspended at the focal plane of the parabolic dish which is at $4.5m$
    from the dish vertex. The measured return loss is much more complex than what
    is found from simulation. 
    While the simulated return loss of the feed structure shows
    two resonant peaks around 120 and 160~MHz, the feed resonance occur at
    slightly higher frequencies. Notable are the difference in the shape of the
    resonant peaks. While the simulated response shows very wide resonant peaks, as
    expected from these dipole structures, the measured return loss has
    a narrow peak at its low frequency resonance. The dominant term in the expression of  return loss of the HERA element in
    receiving mode (equation \ref{eq11}) is the zero delay term $(1+\Gamma_{f})$. Hence the feed return loss and especially the shape of the low frequency resonant peak dominates the overall shape of the delay spectrum in figure
    \ref{fig:sim_em}. 
 The simulated return loss of the feed shows smooth variation across the band and wide resonant peaks resulting in a narrow delay spectrum.

   \subsubsection{CST Time domain simulation}  
    We also compare our measurements with the time domain simulation presented in ~\cite{Ewall-Wice_et_al2016} in figure \ref{fig:sim_em}. In this simulation, the transient electromagnetic response of the HERA element is determined as a function of time when it is subjected to a plane wavefront. The commercial numerical simulation software Microwave Studio, developed by Computer Simulation Technology is used for the simulation. The simulation assumed a constant 125~$\Omega$ impedance at the dipole terminal. Therefore, delay spectrum estimated from this simulation only has the effects of chromaticity introduced due to the structural reflections from the antenna and did not include the mode antenna mode scattering which is a function of frequency if the terminal impedance is frequency dependent. In practice, dipole impedance is function of frequency that results in variation of return loss with frequency. This results in deviation between the simulated delay response and the delay spectrum estimate of our measurements. The simulation and measurements both agree in one aspect. Our measurement also confirms that reflections from the feed structure dominates at lower delays and this is neither an attribute of the reflector dish response nor it is a computational artifact such as windowing. \\
     \subsubsection{Foreground simulation}  
     \label{Foreground_simulation}
    We also compare our measurement with the foreground simulation of~\cite{Thyagarajan_et_al2016} as shown by the shaded regions in the first panel of figure \ref{ds_full_sub_band}. The shaded region of each panel shows the minimum EoR to
    foreground power ratio required for detection of the 21~cm power spectrum at those
    individual delays (or in $k_{\parallel}$ modes). This ratio is computed by
    using a achromatic antenna beam and the sky models. Diffused foreground sky model is
    incorporated from~\cite{deolivieracosta_et_al2008} whereas the point source
    contribution is estimated by combining the NRAO VLA Sky Survey (NVSS) at
    1.42~GHz~\citep{Condon_1998}, Sydney University Molonglo sky Survey (SUMSS)
    ~\citep{Bock_et_al_1999, Mauch_et_al_2003} at 843~MHz. The EoR signal
    is simulated by using 21cmFAST \citep{Messinger_et_al2011}. The model
    parameters used for the EoR signal simulation
    are : Virial temperature of minimum mass of dark matter halos that host
    ionizing sources, $T_{vir}^{min} = 2\times10^4$~K, Ionising efficiency $\eta = 20\%$,
    mean free path of UV photons $R_{mfp} = 15$~Mpc For these model parameter
    values, the redshift of $50\%$ reionization is predicted to be at $z = 8.5$ at
    150~MHz.

    The EoR and foreground power spectra are shown by the blue and the orange curve in figure \ref{fig:cov_weight}. The details of the power spectrum computation are given in section \ref{revision_foreground_sim}. The computed power spectra are associated with the shaded gray regions in following way. For a given separation in delay, for example, for $\tau = 200$ ns, at delay $t = 200$~ns the foreground power spectrum amplitude should reduce from $10^{15}$ (at delay t=0) to $10^4$~mk$^{2}$ (at delay $t = 200$) i.e, 110~dB attenuation of power spectrum amplitude or roughly 55~dB attenuation of power (visibility). For the same $\tau$, between delay $t=$ 100 to 300~ns, the foreground power spectrum amplitude should be attenuated from $10^{14}$ to $10^4$~mk$^{2}$ which is roughly 100~dB i.e 50~dB attenuation of the foreground power relative to the EoR. At $t = 400$~ns, the foreground should be attenuated from $10^{11}$ to $10^3 $~mk$^{2}$ between $t=$200 to 400~ns which is roughly 80~dB attenuation of the foreground power spectrum amplitude or or 40~dB attenuation of the foreground power relative to the EoR. 
For the separation in delay $\tau = 200$~ns, any two points in delay space will need different attenuation of foreground power with respect to the EoR power and the maximum foreground power attenuation required will be 55~dB for the delays where foreground has the highest magnitude. The maximum required attenuation of the foreground power at various values of $\tau$ is plotted in the grey shaded region for different values of $\tau$. 
    
     The EoR to foreground power ratio shows that for a given
    delay mode, for a particular frequency, a successful power spectrum detection at a lower $k_{\parallel}$ mode requires a higher foreground power attenuation relative to the EoR power compared to a higher $k_{\parallel}$ mode. 
    This simulation provides a baseline for the system performance evaluation with a conservative strategy of pure foreground avoidance without any foreground mitigation strategy. The measurement differs from the simulation in one aspect. Simulation includes the achromatic antenna gain excluding the return loss of of power at the antenna terminal due to frequency dependent antenna impedance. This effect is included in the measurements.
  If included in the model, these limits would provide more conservative estimates of the required instrument response which will be even harder to achieve.
          
    \subsubsection{Delay spectrum of sub bands}
    \label{sub-band}
    In the HERA analysis, the 21~cm power spectrum
    is estimated by discretizing the observed data along the line of sight distance i.e
    in frequency or redshift and in the plane of the sky
    $k$~\citep{liu_et_al2014a}. At particular redshift, the bandwidth of
    observation is so chosen that over the corresponding $\Delta z$, the 21~cm
    signal does not significantly evolve. Typically, the 21~cm brightness
    temperature fluctuation evolves over redshift scale $\Delta
    z>0.5$~\citep{loeb_zaldarriaga2004},~\citep{lidz_et_al2008}. Therefore, the delay
    spectrum is estimated over smaller bandwidths.
    
\begin{figure*}[ht]
    \begin{minipage}[b]{0.5\linewidth}
    \centering
    \includegraphics[angle=0, width=\linewidth]{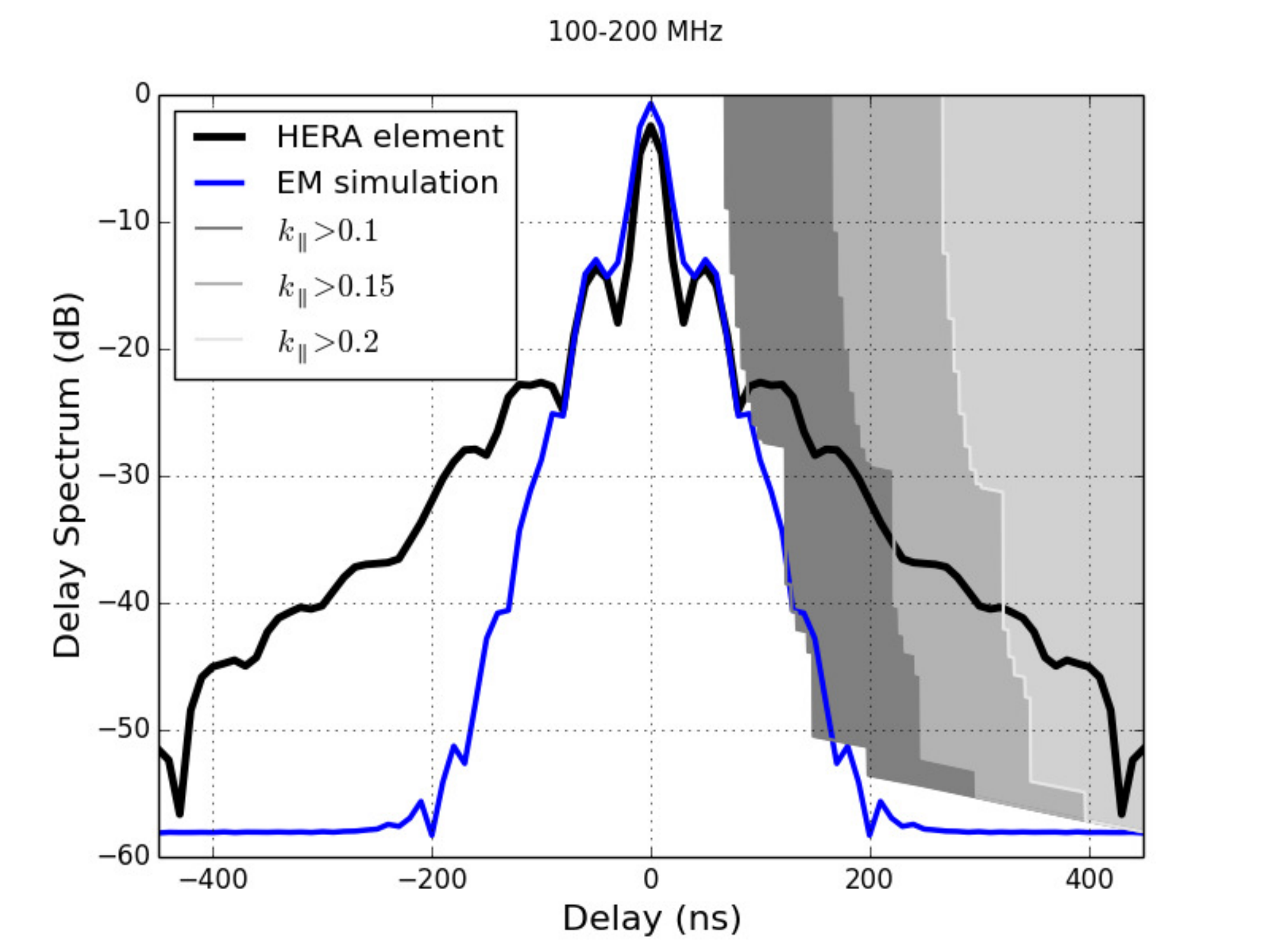}
    \end{minipage}
    \hspace{0.1cm}
    \begin{minipage}[b]{0.5\linewidth}
    \centering
    \includegraphics[angle=0, width=\linewidth]{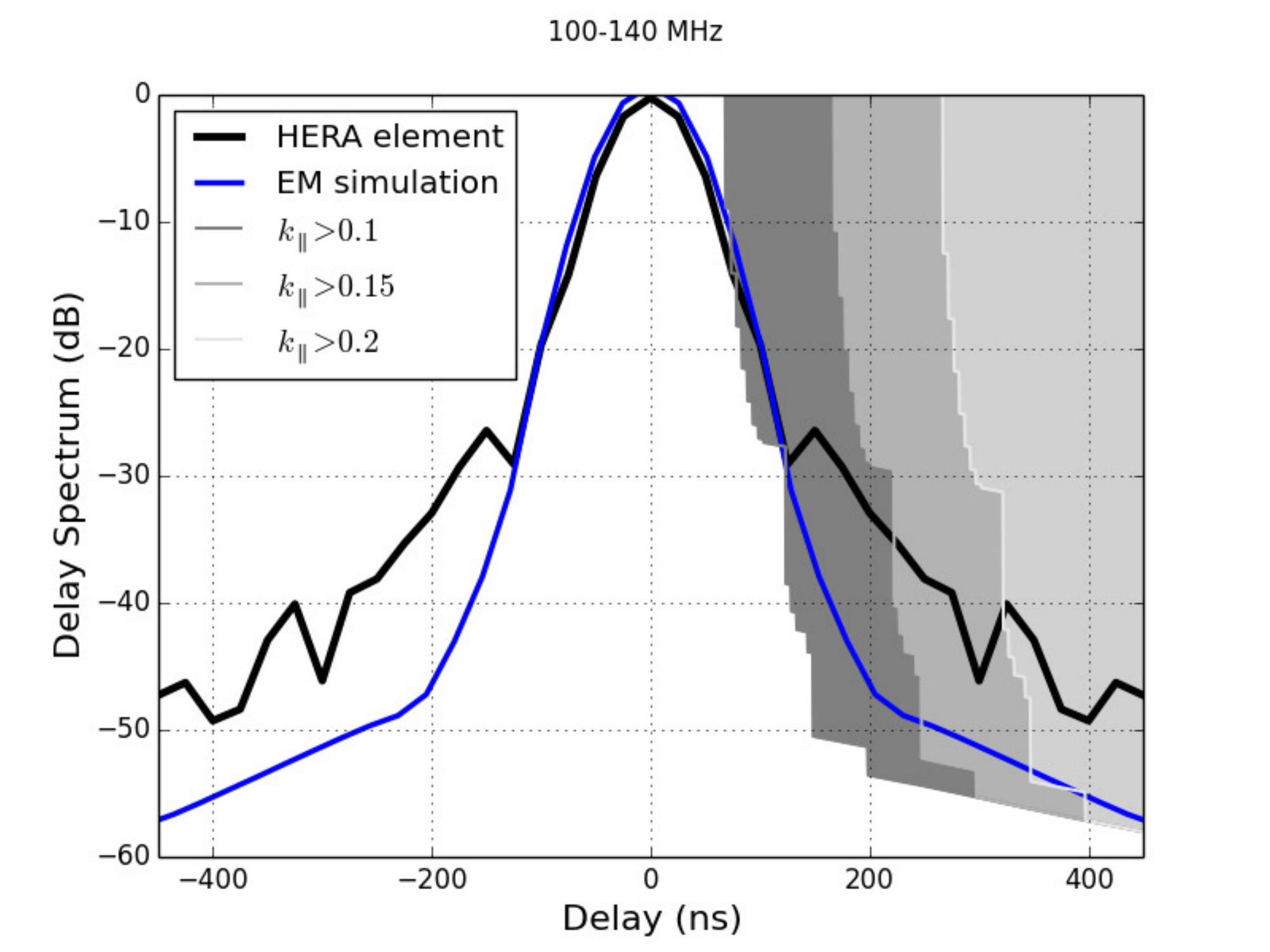}
    \end{minipage}
    \vspace{0.1cm}
    \begin{minipage}[b]{0.5\linewidth}
    \centering
    \includegraphics[angle=0, width=\linewidth]{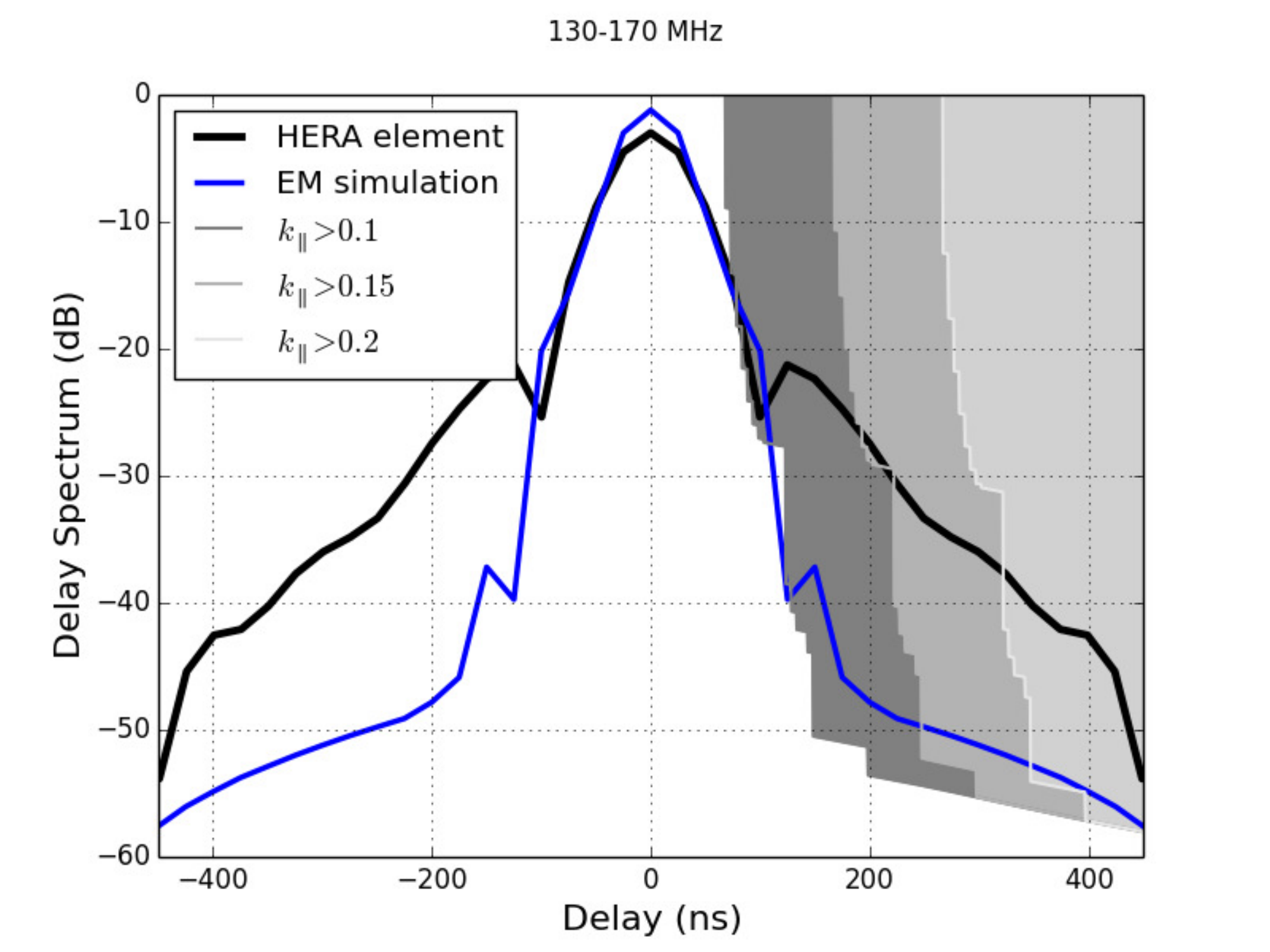}
    \end{minipage}
    \hspace{0.1cm}
    \begin{minipage}[b]{0.5\linewidth}
    \centering
    \includegraphics[angle=0, width=\linewidth]{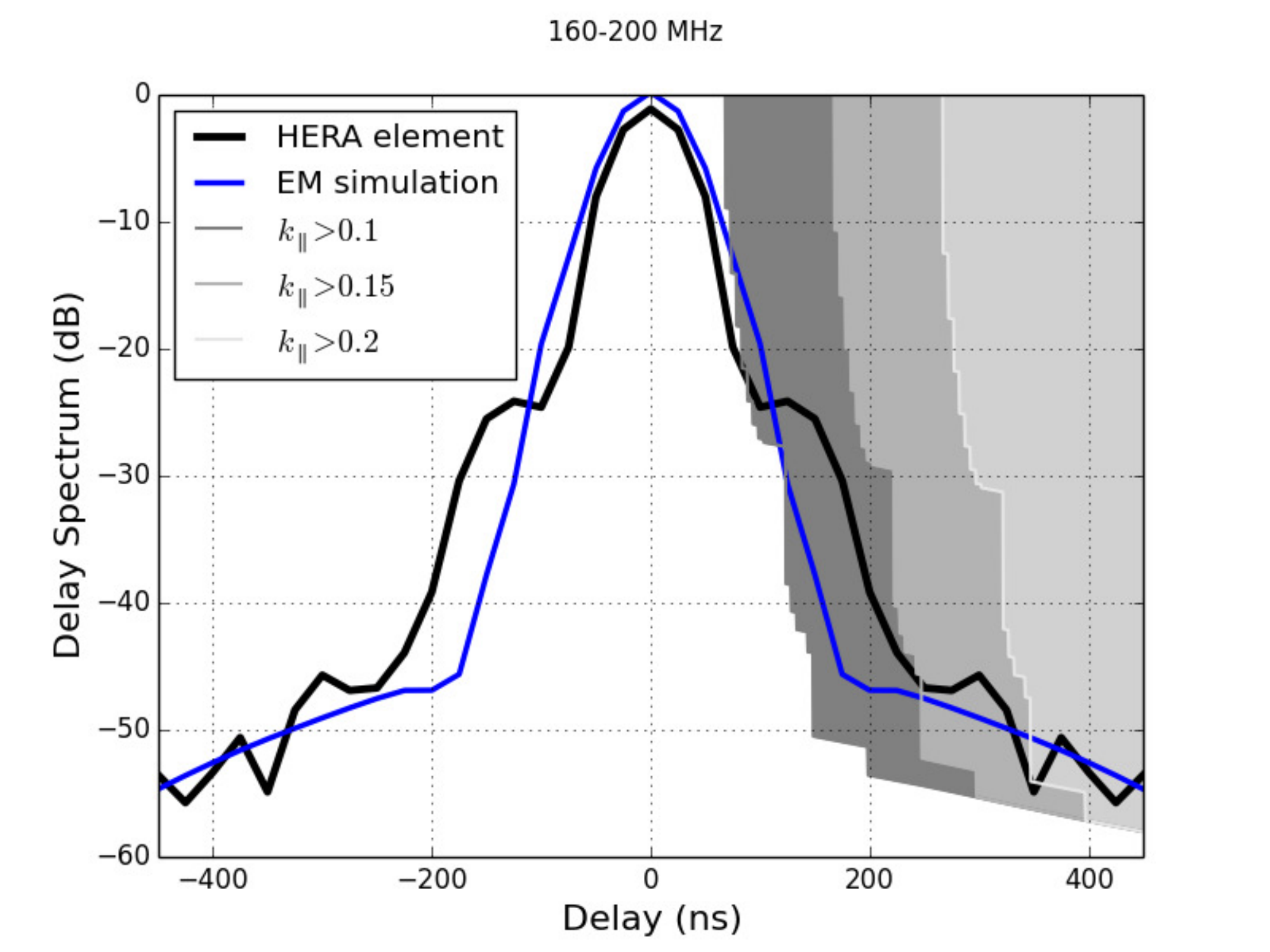}
    \end{minipage}
    \caption{Delay response of the instrument (black) estimated from measurements of the complex return loss of the HERA element. The blue curve shows the same estimated from the EM simulation of the HERA element corrected for the zero delay response. From right to left: the shaded gray regions show the minimum foreground attenuation relative to the EoR power required to detect the EoR power at a given delay (or in corresponding $k_{\parallel}$ mode). At a given delay, detection of EoR power spectrum at smaller $k_{\parallel}$ mode requires higher attenuation of the foreground power. This is shown by different shaded region each represening a minimum $k_{\parallel}$ mode that can be probed for that particular foreground attenuation profile in the delay domain. The instrument delay response estimated over the full band is dominated by the sharpest feature present in the observation band. At $120~MHz$, the feed impedance is best matched to the 50~$\Omega$ reference impedance providing a very low value of return loss. This feature, when Fourier transformed to delay domain results in a wide response.  At higher frequencies , the measured return loss varies smoothly with frequency resulting in narrow delay response of the instrument. }   
    \label{ds_full_sub_band}
    \end{figure*}

    We estimate the instrument
    delay spectrum over various sub bands between 100 to 200~MHz as shown in figure
    \ref{ds_full_sub_band}. Each panel of figure \ref{ds_full_sub_band} shows the
    delay spectrum estimated from the return loss measurement of the HERA element
    between 100-200, 100-140, 130-170, 160 to 200~MHz respectively. Each panel also
    compares the delay spectrum estimate obtained EM simulation of~\cite{deBoer_2016} at various subband. Subband delay spectra are wider compared to full
    band delay spectrum.  The shape of the resonant peaks determines the roll off rate
    of the delay spectrum at all frequency ranges. At higher
    frequency, our measurement closely follows the EM simulations except around
    80-120~ns where the cable reflection between the antenna and the backend manifests
    itself. 
    
    The shaded regions represent the EoR to Foreground power ratio simulated across the entire band.
    The HFSS simulation of chromatic antenna beam varies smoothly across frequency resulting in no significant changes in this simulation at various sub bands.  
     HERA system performance conforms directly to the limits as posed by simulations, at frequencies $>$160~MHz. With the measured system response including the multiple reflections, the instrument would be able to probe the spatial modes of interest $k_{\parallel}>0.2 h$~Mpc$^{-1}$ with very little foreground bleeding around $k_{\parallel} \approx 0.2 h$ Mpc$^{-1}$.
      At other frequencies, it is possible to do a successful measurement by weighting down the
      foreground contribution in the measured data. This is discussed in the following section.
    
 The HERA analysis pipeline for power spectrum estimation exploits the inverse covariance weighting to reduce the foreground contribution to the measured visibility data. In the light of this analysis technique we further investigate the limits of EoR to foreground power ratio from what is established by conservative estimates of \cite{Thyagarajan_et_al2016}. In the following section we briefly describe the power spectrum estimation technique used in HERA analysis and inverse covariance weighting. We evaluate the instrument delay response in the context of covariance weighting.
 
 \section{\textbf{Revised delay spectrum specification using inverse covariance weighting}}    
   \label{revision_foreground_sim}
\begin{figure}
    \centering
    \includegraphics[width=\linewidth]{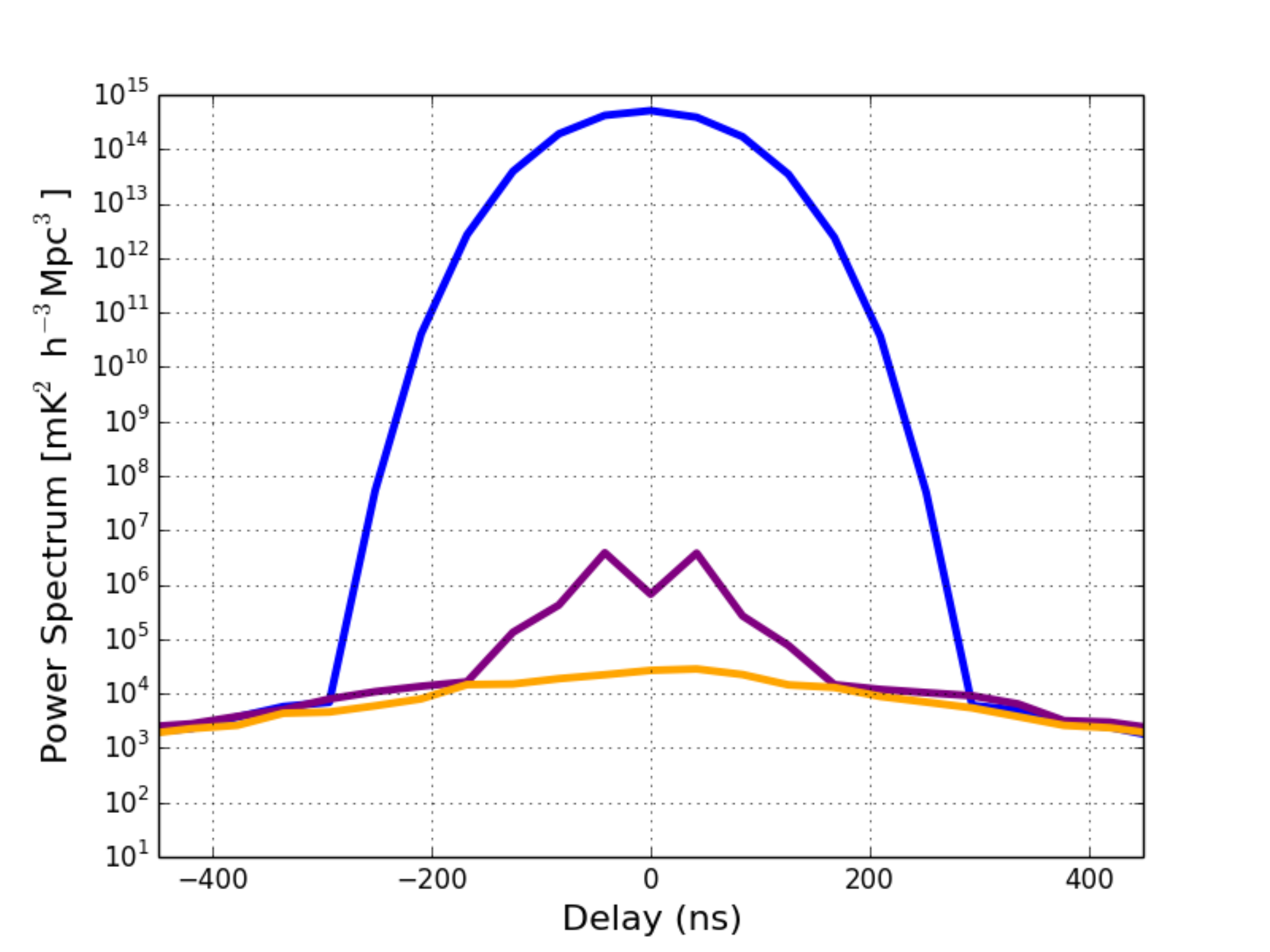}
    \caption{Blue: Power spectrum of the sky signal estimated from the simulated visibility with two adjacent HERA elements and the foreground, EoR model described in section \ref{Foreground_simulation}. Orange: Power spectrum of only the EoR signal. Purple: Power spectrum estimate of the sky signal after weighting the simulated visibility by the inverse of the covariance matrix computed from the simulated data.}
    \label{fig:cov_weight}
    \end{figure}

    Results presented in previous sections focused on the delay-domain performance of
the prototype HERA element which convolves with the delay spectrum of the sky signal and produce the delay spectrum. In the delay domain, after convolution with the instrument response, the foreground response must fall below the amplitude of the 
21cm EoR signal to avoid a systematic bias. 
This simple analysis omits the 
suppression of foreground systematics that are a feature of more sophisticated power spectrum estimation techniques.
In this section, we re-evaluate the specification for the delay-domain performance of the HERA element in light of the HERA power spectrum estimation method using the optimal quadratic estimator (OQE). The OQE formalism has been outlined in extensive detail in  ~\cite{dillon_et_al2013a}, ~\cite{liu_et_al2014a}, ~\cite{liu_et_al2014b}, ~\cite{trott_et_al2012}, ~\cite{Ali_et_al2015}. In order to determine the effect of covariance in the measured data, we notationally describe the OQE formalism here. The 21~cm power spectrum $P_{21}(k_\perp,k_\parallel)$  integrated over a range of $k_\parallel, k_\perp$, is estimated as the band power $\hat p_{\alpha}$, where $\alpha$ represents a range of $k_{\parallel}, k_{\perp}$. The major steps of the OQE formalism are,     
    
    
  The unnormalized
band power $\hat q_\alpha$, estimated from the data vector $\bold x$ of measured visibilities is
    \begin{equation}
    \hat q_{\alpha} = \bold x^{T} E^{\alpha} \bold x, 
    \label{eq15}
    \end{equation}
 $E^{\alpha}$ is a symmetric matrix operation denoting the Fourier transform of the data, binning, and foreground reduction.
The normalized estimate of the power spectrum is,  
    \begin{equation}
    \hat p_{\alpha} = M \hat q_{\alpha} 
    \label{eq17}
    \end{equation} 
    where $M$ is the normalization matrix.
    The true power spectrum $p_{\alpha}$ is  
    \begin{equation}
    \hat p_{\alpha} = W p_{\alpha}
    \label{eq18}
    \end{equation} 
   
   where $W$ is the window function matrix,
 
 Various binning and foreground reduction techniques result in different forms of $E^{\alpha}$ resulting in estimates of $\hat p_{\alpha}$ with different statistics. A possible choice of $E^{\alpha}$ is
    \begin{equation}
    E^{\alpha} =  {1 \over 2} C^{-1} Q_{\alpha} C^{-1}
    \label{eq16}
    \end{equation} 
    
    where $C = \langle \bold x \bold x^{t} \rangle$ is the covariance matrix of the data vector $\bold x$. $Q_{\alpha}$ is a matrix operator that Fourier transforms the visibilities along the frequency axis and maps them into the $\bold k$ space. 
 The critical step in the OQE formalism for power spectrum estimation is the weighting of the data by the inverse of its covariance. This can result in orders-of-magnitude reduction in the foreground power relative to the EoR signal.
To derive a delay-spectrum specification for a prototype HERA element using the inverse covariance weighting, we begin
with the
   simulations of \cite{Thyagarajan_et_al2016} and use OQE formalism to estimate the amplitude of the power spectrum
from a simulated data vector $\bold x$ 
    which contains contributions from the EoR, the foreground and
   the instrument noise. The covariance matrix is estimated from 80 different realizations of the data vector $\bold x$ between $0-24$~h of LST. The power spectrum, estimated using one realization of $\bold x$, with and without covariance weighting are shown in figure \ref{fig:cov_weight}. The orange curve represents the power spectrum computed from the EoR model for two adjacent HERA elements with a given baseline orientation. The blue curve shows the power spectrum of the observed sky signal including both foreground and the EoR using the delay transform technique after appropriate windowing. Here, the contribution from the bright foregrounds dominate the lower delay modes, making separation of the EoR from the foreground impossible up to delays $\approx 300$~ns. 
The power spectrum estimated after weighting the data by the inverse of the covariance matrix is illustrated by the purple curve in the plot. Weighting the simulated data by the covariance matrix which is also estimated from the simulated visibility dramatically reduces the foreground power relative to the EoR signal at low delays, opening up the possibility of estimating the EoR power spectrum at those delays. 

\begin{figure}
    \centering
    \includegraphics[width=\linewidth]{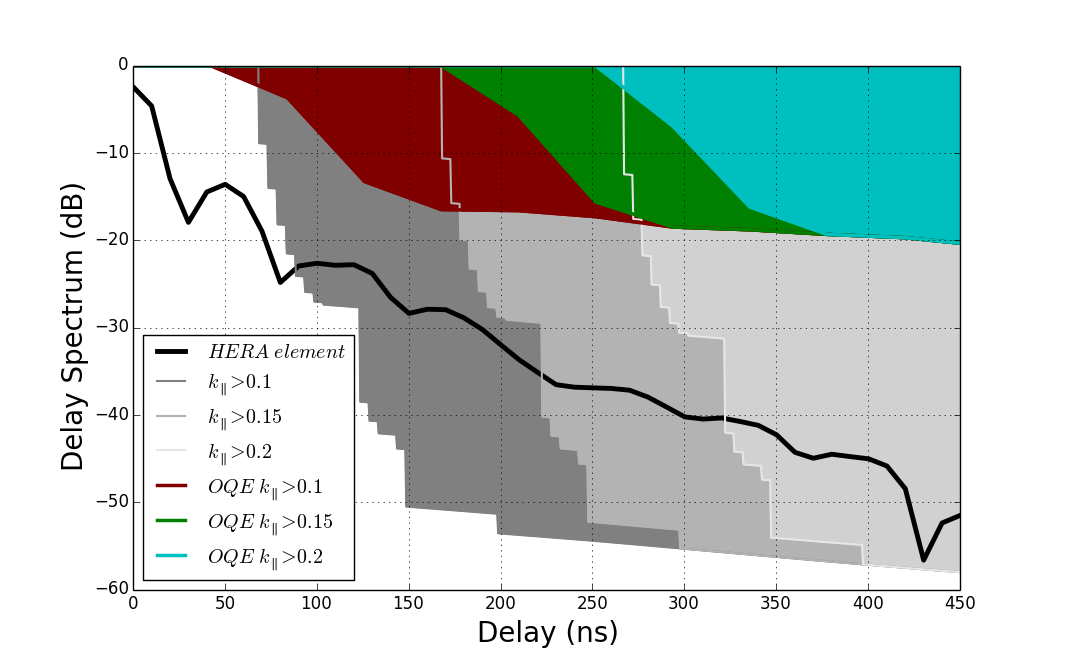}
    \caption{Delay spectrum of HERA element estimated from the reflectometry measurements by a vector network analyzer (black). The colored region shows the revised EoR to foreground power ratio as a function of the separation in delay space after weighting the simulated visibility by the inverse of its covariance. Due to inverse covariance weighting of the visibility, the foreground contribution at each delay is reduced relaxing the limits on the required foreground attenuation relative to the EoR by about $30$~dB at each delay. In comparison to this revised specification derived from the sky power simulation, the HERA element delay response demonstrates its capacity for a successful power spectrum detection without any additional design improvement.}
    \label{fig:sim_fg_revised}
    \end{figure}

In a simple delay-transform power spectrum estimation, the chromatic antenna response is convolved with the foreground signal in delay domain.  As described in \cite{Thyagarajan_et_al2016}, this 
enables us to set a specification
for the reflectometry of the HERA element as a function of delay, as illustrated by the grey shaded regions in
Figure \ref{fig:sim_fg_revised}.  This relationship is not so straightforward when using
optimal quadratic power-spectrum estimation.  For example, a direction-independent bandpass shape that multiplies the foreground signal falls into a single delay mode that can be downweighted essentially to zero.
This makes it problematic to interpret the output power spectrum in the light of reflectometry constraints which comes from the beam integrated reflectometry measurements. Without
knowing in detail the direction dependence of the HERA element chromaticity, it is impossible to estimate exactly how many Eigen modes
will be occupied by the systematics arising from the foreground-HERA element interaction. 

In a relatively conservative estimate, we simply use the inverse covariance weighted foreground power spectrum from simulations as an effective input foreground amplitude and repeat the translation to a reflectometry specification as before.  The result is illustrated by the colored shaded regions
in Figure \ref{fig:sim_fg_revised}. This approach ignores the ability of OQE to identify and invert
instrumental covariances, making it a relatively conservative standard.  On the other hand, should the number
of direction-dependent spectral eigenmodes of the dish become large, this approach could potentially underestimate
the impact of dish chromaticity.  Therefore, we present it as a demonstration that our reflectometry specifications
could be substantially less stringent for HERA's OQE pipeline, but suggest not over-interpreting the exact level of the implied specification without a detailed analysis of the direction dependence of the reflectometry
results.

The reflectometry measurements demonstrate the performance of the HERA element including it intrinsic chromaticity and chromaticity generated due to multiple reflections in the system. The prototype HERA element delay response, in conjunction with the inverse covariance method of foreground suppression, indicates that the HERA prototype element satisfies the necessary condition to make a successful power spectrum measurement for the spatial modes as low as $k_{\parallel} >0.1 h$ Mpc$^{-1}$.
    \section{\textbf{Conclusion}}
    The interplay between the extremely bright sky signal and the system response
    has remained somewhat undetermined for the first generation 21cm experiments such
    as MWA, LOFAR, PAPER. With the theory of redshifted 21cm experiments very
    well evolved, it is now absolutely necessary to quantify the instrumental limits
    of these measurements in order to produce accurate power spectrum estimates.
    The delay-domain
    performance of the HERA dish is central to HERA's function as a power spectrum
    instrument. In this paper, we studied the performance of a prototype HERA element in both frequency
    as well as delay domain. We introduced a mathematical formalism that explicitly
    relate the delay response of the HERA element to EoR to foreground power ratio in the delay domain.
    The effects of multiple reflections  in a HERA
    element is investigated in detail and their effects on the measured
    visibility is estimated. Reflectometry measurements characterized HERA's performance in delay domain
    domain and, as equation \ref{eq10} shows, these measurements must be adjusted
    for a difference in transmission/reception at the first feed encounter in
    order to be interpreted as the delay response of a HERA element relative to an incident
    plane wave from the sky.  It is also shown that the windowing
    the measured data by the Blackman-Harris window, shown in \cite{Thyagarajan_et_al2016} is critical for accurately measuring the antenna delay response at
    higher delays, where sidelobes from much higher amplitude responses at small
    delays can easily dominate.  Windowing the measured data conclusively shows that the lower delay response results from the structural reflections in the feed. Given the critical nature
    of the windowing function, it is recommended that all reflectometry
    measurements be performed in the frequency domain, so that the data could be
    Fourier transformed with the appropriate window.  The delay spectrum estimates
    are compared with the electromagnetic simulation of
    the HERA element~\citep{ddboer_et_al2016, Ewall-Wice_et_al2016}. The
    same is then compared with the estimates derived from the foreground simulation
    of~\cite{Thyagarajan_et_al2016}. The performance is also evaluated in the light
    of HERA power spectrum estimation technique using inverse covariance weighting
    formalism.  Taken all together, we conclude that the HERA antenna element,
    with a PAPER-style crossed dipole feed and cylindrical cage, satisfies the criteria necessary to meet its science
    goal. \\
 \textbf{Acknowledgements:}  
This work was supported by the U.S. National Science Foundation (NSF) through awards AST-1440343 $\&$ AST-1410719. ARP acknowledges support from NSF CAREER award 13 52519. AL acknowledges support by NASA through Hubble Fellowship grant \#HST-HF2-51363.001-A awarded by the Space Telescope Science Institute, operated by the Association of Universities for Research in Astronomy, Inc., for NASA, under contract NAS5-26555. This work is completed as part of the University of California Cosmic Dawn Initiative. AL, ARP, and SRF acknowledge support from the UC Office of the President Multicampus Research Programs and Initiatives through award MR-15-328388

    
    \bibliographystyle{apj}
    \bibliography{Reference}{}

\begin{thebibliography}{}
\expandafter\ifx\csname natexlab\endcsname\relax\def\natexlab#1{#1}\fi

\bibitem[{{Ali} {et~al.}(2015){Ali}, {Parsons}, {Zheng}, {Pober}, {Liu},
  {Aguirre}, {Bradley}, {Bernardi}, {Carilli}, {Cheng}, {DeBoer}, {Dexter},
  {Grobbelaar}, {Horrell}, {Jacobs}, {Klima}, {MacMahon}, {Maree}, {Moore},
  {Razavi}, {Stefan}, {Walbrugh}, \& {Walker}}]{Ali_et_al2015}
{Ali}, Z.~S., {Parsons}, A.~R., {Zheng}, H., {et~al.} 2015, \apj, 809, 61

\bibitem[{{Barry} {et~al.}(2016){Barry}, {Hazelton}, {Sullivan}, {Morales}, \&
  {Pober}}]{2016MNRAS.461.3135B}
{Barry}, N., {Hazelton}, B., {Sullivan}, I., {Morales}, M.~F., \& {Pober},
  J.~C. 2016, \mnras, 461, 3135

\bibitem[{{Bock} {et~al.}(1999){Bock}, {Large}, \& {Sadler}}]{Bock_et_al_1999}
{Bock}, D.~C.-J., {Large}, M.~I., \& {Sadler}, E.~M. 1999, \aj, 117, 1578

\bibitem[{{Bowman} {et~al.}(2009){Bowman}, {Morales}, \&
  {Hewitt}}]{bowman_et_al2009}
{Bowman}, J.~D., {Morales}, M.~F., \& {Hewitt}, J.~N. 2009, \apj, 695, 183

\bibitem[{{Bowman} \& {Rogers}(2010)}]{Bowman_et_al2010}
{Bowman}, J.~D., \& {Rogers}, A.~E.~E. 2010, \nat, 468, 796

\bibitem[{{Condon} {et~al.}(1998){Condon}, {Cotton}, {Greisen}, {Yin},
  {Perley}, {Taylor}, \& {Broderick}}]{Condon_1998}
{Condon}, J.~J., {Cotton}, W.~D., {Greisen}, E.~W., {et~al.} 1998, \aj, 115,
  1693

\bibitem[{{Datta} {et~al.}(2010){Datta}, {Bowman}, \&
  {Carilli}}]{datta_et_al2010}
{Datta}, A., {Bowman}, J.~D., \& {Carilli}, C.~L. 2010, \apj, 724, 526

\bibitem[{{de Oliveira-Costa} {et~al.}(2008){de Oliveira-Costa}, {Tegmark},
  {Gaensler}, {Jonas}, {Landecker}, \& {Reich}}]{deolivieracosta_et_al2008}
{de Oliveira-Costa}, A., {Tegmark}, M., {Gaensler}, B.~M., {et~al.} 2008,
  \mnras, 388, 247

\bibitem[{{DeBoer} {et~al.}(2016{\natexlab{a}}){DeBoer}, {Parsons}, {Aguirre},
  {Alexander}, {Ali}, {Beardsley}, {Bernardi}, {Bowman}, {Bradley}, {Carilli},
  {Cheng}, {de Lera Acedo}, {Dillon}, {Ewall-Wice}, {Fadana}, {Fagnoni},
  {Fritz}, {Furlanetto}, {Glendenning}, {Greig}, {Grobbelaar}, {Hazelton},
  {Hewitt}, {Hickish}, {Jacobs}, {Julius}, {Kariseb}, {Kohn}, {Lekalake},
  {Liu}, {Loots}, {MacMahon}, {Malan}, {Malgas}, {Maree}, {Mathison},
  {Matsetela}, {Mesinger}, {Morales}, {Neben}, {Patra}, {Pieterse}, {Pober},
  {Razavi-Ghods}, {Ringuette}, {Robnett}, {Rosie}, {Sell}, {Smith}, {Syce},
  {Tegmark}, {Thyagarajan}, {Williams}, \& {Zheng}}]{deBoer_2016}
{DeBoer}, D.~R., {Parsons}, A.~R., {Aguirre}, J.~E., {et~al.}
  2016{\natexlab{a}}, ArXiv e-prints, arXiv:1606.07473

\bibitem[{{DeBoer} {et~al.}(2016{\natexlab{b}}){DeBoer}, {Parsons}, {Aguirre},
  {Alexander}, {Ali}, {Beardsley}, {Bernardi}, {Bowman}, {Bradley}, {Carilli},
  {Cheng}, {de Lera Acedo}, {Dillon}, {Ewall-Wice}, {Fadana}, {Fagnoni},
  {Fritz}, {Furlanetto}, {Glendenning}, {Greig}, {Grobbelaar}, {Hazelton},
  {Hewitt}, {Hickish}, {Jacobs}, {Julius}, {Kariseb}, {Kohn}, {Lekalake},
  {Liu}, {Loots}, {MacMahon}, {Malan}, {Malgas}, {Maree}, {Mathison},
  {Matsetela}, {Mesinger}, {Morales}, {Neben}, {Patra}, {Pieterse}, {Pober},
  {Razavi-Ghods}, {Ringuette}, {Robnett}, {Rosie}, {Sell}, {Smith}, {Syce},
  {Tegmark}, {Thyagarajan}, {Williams}, \& {Zheng}}]{ddboer_et_al2016}
---. 2016{\natexlab{b}}, ArXiv e-prints, arXiv:1606.07473

\bibitem[{{Dillon} {et~al.}(2013){Dillon}, {Liu}, \&
  {Tegmark}}]{dillon_et_al2013a}
{Dillon}, J.~S., {Liu}, A., \& {Tegmark}, M. 2013, \prd, 87, 043005

\bibitem[{{Dillon} {et~al.}(2014){Dillon}, {Liu}, {Williams}, {Hewitt},
  {Tegmark}, {Morgan}, {Levine}, {Morales}, {Tingay}, {Bernardi}, {Bowman},
  {Briggs}, {Cappallo}, {Emrich}, {Mitchell}, {Oberoi}, {Prabu}, {Wayth}, \&
  {Webster}}]{2014PhRvD..89b3002D}
{Dillon}, J.~S., {Liu}, A., {Williams}, C.~L., {et~al.} 2014, \prd, 89, 023002

\bibitem[{{Dillon} {et~al.}(2015){Dillon}, {Tegmark}, {Liu}, {Ewall-Wice},
  {Hewitt}, {Morales}, {Neben}, {Parsons}, \& {Zheng}}]{2015PhRvD..91b3002D}
{Dillon}, J.~S., {Tegmark}, M., {Liu}, A., {et~al.} 2015, \prd, 91, 023002

\bibitem[{{Ewall-Wice} {et~al.}(2016){Ewall-Wice}, {Bradley}, {DeBoer},
  {Hewitt}, {Parsons}, {Aguirre}, {Ali}, {Bowman}, {Cheng}, {Neben}, {Patra},
  {Thyagarajan}, {Venter}, {de Lera Acedo}, {Dillon}, {Doolittle}, {Egan},
  {Hendrick}, {Klima}, {Kohn}, {Schaffner}, {Shelton}, {Saliwanchik},
  {Tegmark}, {Taylor}, {Taylor}, \& {Wirt}}]{Ewall-Wice_et_al2016}
{Ewall-Wice}, A., {Bradley}, R., {DeBoer}, D., {et~al.} 2016, ArXiv e-prints,
  arXiv:1602.06277

\bibitem[{{Furlanetto} {et~al.}(2006){Furlanetto}, {Oh}, \&
  {Briggs}}]{furlanetto_et_al2006}
{Furlanetto}, S.~R., {Oh}, S.~P., \& {Briggs}, F.~H. 2006, \physrep, 433, 181

\bibitem[{{Kohn} {et~al.}(2016){Kohn}, {Aguirre}, {Nunhokee}, {Bernardi},
  {Pober}, {Ali}, {Bradley}, {Carilli}, {DeBoer}, {Gugliucci}, {Jacobs},
  {Klima}, {MacMahon}, {Manley}, {Moore}, {Parsons}, {Stefan}, \&
  {Walbrugh}}]{Saul_et_al2016}
{Kohn}, S.~A., {Aguirre}, J.~E., {Nunhokee}, C.~D., {et~al.} 2016, \apj, 823,
  88

\bibitem[{{Lidz} {et~al.}(2008){Lidz}, {Zahn}, {McQuinn}, {Zaldarriaga}, \&
  {Hernquist}}]{lidz_et_al2008}
{Lidz}, A., {Zahn}, O., {McQuinn}, M., {Zaldarriaga}, M., \& {Hernquist}, L.
  2008, \apj, 680, 962

\bibitem[{Liu {et~al.}(2014{\natexlab{a}})Liu, Parsons, \&
  Trott}]{liu_et_al2014a}
Liu, A., Parsons, A.~R., \& Trott, C.~M. 2014{\natexlab{a}}, Phys. Rev. D, 90,
  023018

\bibitem[{Liu {et~al.}(2014{\natexlab{b}})Liu, Parsons, \&
  Trott}]{liu_et_al2014b}
---. 2014{\natexlab{b}}, Phys. Rev. D, 90, 023019

\bibitem[{{Loeb} \& {Zaldarriaga}(2004)}]{loeb_zaldarriaga2004}
{Loeb}, A., \& {Zaldarriaga}, M. 2004, Physical Review Letters, 92, 211301

\bibitem[{{Madau} {et~al.}(1997){Madau}, {Meiksin}, \&
  {Rees}}]{1997ApJ...475..429M}
{Madau}, P., {Meiksin}, A., \& {Rees}, M.~J. 1997, \apj, 475, 429

\bibitem[{{Mauch} {et~al.}(2003){Mauch}, {Murphy}, {Buttery}, {Curran},
  {Hunstead}, {Piestrzynski}, {Robertson}, \& {Sadler}}]{Mauch_et_al_2003}
{Mauch}, T., {Murphy}, T., {Buttery}, H.~J., {et~al.} 2003, \mnras, 342, 1117

\bibitem[{{Mellema} {et~al.}(2013){Mellema}, {Koopmans}, {Abdalla}, {Bernardi},
  {Ciardi}, {Daiboo}, {de Bruyn}, {Datta}, {Falcke}, {Ferrara}, {Iliev},
  {Iocco}, {Jeli{\'c}}, {Jensen}, {Joseph}, {Labroupoulos}, {Meiksin},
  {Mesinger}, {Offringa}, {Pandey}, {Pritchard}, {Santos}, {Schwarz},
  {Semelin}, {Vedantham}, {Yatawatta}, \& {Zaroubi}}]{2013ExA....36..235M}
{Mellema}, G., {Koopmans}, L.~V.~E., {Abdalla}, F.~A., {et~al.} 2013,
  Experimental Astronomy, 36, 235

\bibitem[{{Mesinger} {et~al.}(2011){Mesinger}, {Furlanetto}, \&
  {Cen}}]{Messinger_et_al2011}
{Mesinger}, A., {Furlanetto}, S., \& {Cen}, R. 2011, \mnras, 411, 955

\bibitem[{{Morales} {et~al.}(2012){Morales}, {Hazelton}, {Sullivan}, \&
  {Beardsley}}]{2012ApJ...752..137M}
{Morales}, M.~F., {Hazelton}, B., {Sullivan}, I., \& {Beardsley}, A. 2012,
  \apj, 752, 137

\bibitem[{{Morales} \& {Wyithe}(2010)}]{morales_wyithe2010}
{Morales}, M.~F., \& {Wyithe}, J.~S.~B. 2010, \araa, 48, 127

\bibitem[{{Neben} {et~al.}(2016){Neben}, {Bradley}, {Hewitt}, {DeBoer},
  {Parsons}, {Aguirre}, {Ali}, {Cheng}, {Ewall-Wice}, {Patra}, {Thyagarajan},
  {Bowman}, {Dickenson}, {Dillon}, {Doolittle}, {Egan}, {Hedrick}, {Jacobs},
  {Kohn}, {Klima}, {Moodley}, {Saliwanchik}, {Schaffner}, {Shelton}, {Taylor},
  {Taylor}, {Tegmark}, {Wirt}, \& {Zheng}}]{Neben_et_al2016}
{Neben}, A.~R., {Bradley}, R.~F., {Hewitt}, J.~N., {et~al.} 2016, \apj, 826,
  199

\bibitem[{{Paciga} {et~al.}(2011){Paciga}, {Chang}, {Gupta}, {Nityanada},
  {Odegova}, {Pen}, {Peterson}, {Roy}, \& {Sigurdson}}]{Paciga_et_al2011}
{Paciga}, G., {Chang}, T.-C., {Gupta}, Y., {et~al.} 2011, \mnras, 413, 1174

\bibitem[{{Parsons} {et~al.}(2012{\natexlab{a}}){Parsons}, {Pober}, {McQuinn},
  {Jacobs}, \& {Aguirre}}]{parsons_et_al2012a}
{Parsons}, A., {Pober}, J., {McQuinn}, M., {Jacobs}, D., \& {Aguirre}, J.
  2012{\natexlab{a}}, \apj, 753, 81

\bibitem[{{Parsons} \& {Backer}(2009)}]{Parsons_Backer_2009}
{Parsons}, A.~R., \& {Backer}, D.~C. 2009, \aj, 138, 219

\bibitem[{{Parsons} {et~al.}(2012{\natexlab{b}}){Parsons}, {Pober}, {Aguirre},
  {Carilli}, {Jacobs}, \& {Moore}}]{Parsons_et_al_2012}
{Parsons}, A.~R., {Pober}, J.~C., {Aguirre}, J.~E., {et~al.}
  2012{\natexlab{b}}, \apj, 756, 165

\bibitem[{{Parsons} {et~al.}(2010){Parsons}, {Backer}, {Foster}, {Wright},
  {Bradley}, {Gugliucci}, {Parashare}, {Benoit}, {Aguirre}, {Jacobs},
  {Carilli}, {Herne}, {Lynch}, {Manley}, \& {Werthimer}}]{parsons_et_al2010}
{Parsons}, A.~R., {Backer}, D.~C., {Foster}, G.~S., {et~al.} 2010, \aj, 139,
  1468

\bibitem[{{Parsons} {et~al.}(2014){Parsons}, {Liu}, {Aguirre}, {Ali},
  {Bradley}, {Carilli}, {DeBoer}, {Dexter}, {Gugliucci}, {Jacobs}, {Klima},
  {MacMahon}, {Manley}, {Moore}, {Pober}, {Stefan}, \&
  {Walbrugh}}]{parsons_et_al2014}
{Parsons}, A.~R., {Liu}, A., {Aguirre}, J.~E., {et~al.} 2014, \apj, 788, 106

\bibitem[{{Patra} {et~al.}(2015{\natexlab{a}}){Patra}, {Bray}, {Ekers}, \&
  {Roberts}}]{2015arXiv150205862P}
{Patra}, N., {Bray}, J., {Ekers}, R., \& {Roberts}, P. 2015{\natexlab{a}},
  ArXiv e-prints, arXiv:1502.05862

\bibitem[{{Patra} {et~al.}(2013){Patra}, {Subrahmanyan}, {Raghunathan}, \&
  {Udaya Shankar}}]{Patra_et_al2013}
{Patra}, N., {Subrahmanyan}, R., {Raghunathan}, A., \& {Udaya Shankar}, N.
  2013, Experimental Astronomy, 36, 319

\bibitem[{{Patra} {et~al.}(2015{\natexlab{b}}){Patra}, {Subrahmanyan}, {Sethi},
  {Udaya Shankar}, \& {Raghunathan}}]{Patra_et_al2015}
{Patra}, N., {Subrahmanyan}, R., {Sethi}, S., {Udaya Shankar}, N., \&
  {Raghunathan}, A. 2015{\natexlab{b}}, \apj, 801, 138

\bibitem[{{Pober} {et~al.}(2013){Pober}, {Parsons}, {Aguirre}, {Ali},
  {Bradley}, {Carilli}, {DeBoer}, {Dexter}, {Gugliucci}, {Jacobs}, {Klima},
  {MacMahon}, {Manley}, {Moore}, {Stefan}, \& {Walbrugh}}]{pober_et_al2013}
{Pober}, J.~C., {Parsons}, A.~R., {Aguirre}, J.~E., {et~al.} 2013, \apjl, 768,
  L36

\bibitem[{{Presley} {et~al.}(2015){Presley}, {Liu}, \&
  {Parsons}}]{presley_et_al2015}
{Presley}, M., {Liu}, A., \& {Parsons}, A. 2015, ArXiv e-prints: 1501.01633,
  arXiv:1501.01633

\bibitem[{{Pritchard} \& {Loeb}(2012)}]{pritchard_loeb2012}
{Pritchard}, J.~R., \& {Loeb}, A. 2012, Reports on Progress in Physics, 75,
  086901

\bibitem[{{Shaver} {et~al.}(1999){Shaver}, {Windhorst}, {Madau}, \& {de
  Bruyn}}]{Shaver_et_al1999}
{Shaver}, P.~A., {Windhorst}, R.~A., {Madau}, P., \& {de Bruyn}, A.~G. 1999,
  \aap, 345, 380

\bibitem[{{Thyagarajan} {et~al.}(2016){Thyagarajan}, {Parsons}, {DeBoer},
  {Bowman}, {Ewall-Wice}, {Neben}, \& {Patra}}]{Thyagarajan_et_al2016}
{Thyagarajan}, N., {Parsons}, A.~R., {DeBoer}, D.~R., {et~al.} 2016, \apj, 825,
  9

\bibitem[{{Thyagarajan} {et~al.}(2013){Thyagarajan}, {Udaya Shankar},
  {Subrahmanyan}, {Arcus}, {Bernardi}, {Bowman}, {Briggs}, {Bunton},
  {Cappallo}, {Corey}, {deSouza}, {Emrich}, {Gaensler}, {Goeke}, {Greenhill},
  {Hazelton}, {Herne}, {Hewitt}, {Johnston-Hollitt}, {Kaplan}, {Kasper},
  {Kincaid}, {Koenig}, {Kratzenberg}, {Lonsdale}, {Lynch}, {McWhirter},
  {Mitchell}, {Morales}, {Morgan}, {Oberoi}, {Ord}, {Pathikulangara},
  {Remillard}, {Rogers}, {Anish Roshi}, {Salah}, {Sault}, {Srivani}, {Stevens},
  {Thiagaraj}, {Tingay}, {Wayth}, {Waterson}, {Webster}, {Whitney}, {Williams},
  {Williams}, \& {Wyithe}}]{nithya_et_al2013}
{Thyagarajan}, N., {Udaya Shankar}, N., {Subrahmanyan}, R., {et~al.} 2013,
  \apj, 776, 6

\bibitem[{{Thyagarajan} {et~al.}(2015){Thyagarajan}, {Jacobs}, {Bowman},
  {Barry}, {Beardsley}, {Bernardi}, {Briggs}, {Cappallo}, {Carroll}, {Corey},
  {de Oliveira-Costa}, {Dillon}, {Emrich}, {Ewall-Wice}, {Feng}, {Goeke},
  {Greenhill}, {Hazelton}, {Hewitt}, {Hurley-Walker}, {Johnston-Hollitt},
  {Kaplan}, {Kasper}, {Kim}, {Kittiwisit}, {Kratzenberg}, {Lenc}, {Line},
  {Loeb}, {Lonsdale}, {Lynch}, {McKinley}, {McWhirter}, {Mitchell}, {Morales},
  {Morgan}, {Neben}, {Oberoi}, {Offringa}, {Ord}, {Paul}, {Pindor}, {Pober},
  {Prabu}, {Procopio}, {Riding}, {Rogers}, {Roshi}, {Udaya Shankar}, {Sethi},
  {Srivani}, {Subrahmanyan}, {Sullivan}, {Tegmark}, {Tingay}, {Trott},
  {Waterson}, {Wayth}, {Webster}, {Whitney}, {Williams}, {Williams}, {Wu}, \&
  {Wyithe}}]{Thyagarajan_et_al2015}
{Thyagarajan}, N., {Jacobs}, D.~C., {Bowman}, J.~D., {et~al.} 2015, \apj, 804,
  14

\bibitem[{{Trott} {et~al.}(2012){Trott}, {Wayth}, \&
  {Tingay}}]{trott_et_al2012}
{Trott}, C.~M., {Wayth}, R.~B., \& {Tingay}, S.~J. 2012, \apj, 757, 101

\bibitem[{{van Haarlem} {et~al.}(2013){van Haarlem}, {Wise}, {Gunst}, {Heald},
  {McKean}, {Hessels}, {de Bruyn}, {Nijboer}, {Swinbank}, {Fallows},
  {Brentjens}, {Nelles}, {Beck}, {Falcke}, {Fender}, {H{\"o}randel},
  {Koopmans}, {Mann}, {Miley}, {R{\"o}ttgering}, {Stappers}, {Wijers},
  {Zaroubi}, {van den Akker}, {Alexov}, {Anderson}, {Anderson}, {van Ardenne},
  {Arts}, {Asgekar}, {Avruch}, {Batejat}, {B{\"a}hren}, {Bell}, {Bell}, {van
  Bemmel}, {Bennema}, {Bentum}, {Bernardi}, {Best}, {B{\^i}rzan}, {Bonafede},
  {Boonstra}, {Braun}, {Bregman}, {Breitling}, {van de Brink}, {Broderick},
  {Broekema}, {Brouw}, {Br{\"u}ggen}, {Butcher}, {van Cappellen}, {Ciardi},
  {Coenen}, {Conway}, {Coolen}, {Corstanje}, {Damstra}, {Davies}, {Deller},
  {Dettmar}, {van Diepen}, {Dijkstra}, {Donker}, {Doorduin}, {Dromer}, {Drost},
  {van Duin}, {Eisl{\"o}ffel}, {van Enst}, {Ferrari}, {Frieswijk}, {Gankema},
  {Garrett}, {de Gasperin}, {Gerbers}, {de Geus}, {Grie{\ss}meier}, {Grit},
  {Gruppen}, {Hamaker}, {Hassall}, {Hoeft}, {Holties}, {Horneffer}, {van der
  Horst}, {van Houwelingen}, {Huijgen}, {Iacobelli}, {Intema}, {Jackson},
  {Jelic}, {de Jong}, {Juette}, {Kant}, {Karastergiou}, {Koers}, {Kollen},
  {Kondratiev}, {Kooistra}, {Koopman}, {Koster}, {Kuniyoshi}, {Kramer},
  {Kuper}, {Lambropoulos}, {Law}, {van Leeuwen}, {Lemaitre}, {Loose}, {Maat},
  {Macario}, {Markoff}, {Masters}, {McFadden}, {McKay-Bukowski}, {Meijering},
  {Meulman}, {Mevius}, {Middelberg}, {Millenaar}, {Miller-Jones}, {Mohan},
  {Mol}, {Morawietz}, {Morganti}, {Mulcahy}, {Mulder}, {Munk}, {Nieuwenhuis},
  {van Nieuwpoort}, {Noordam}, {Norden}, {Noutsos}, {Offringa}, {Olofsson},
  {Omar}, {Orr{\'u}}, {Overeem}, {Paas}, {Pandey-Pommier}, {Pandey}, {Pizzo},
  {Polatidis}, {Rafferty}, {Rawlings}, {Reich}, {de Reijer}, {Reitsma},
  {Renting}, {Riemers}, {Rol}, {Romein}, {Roosjen}, {Ruiter}, {Scaife}, {van
  der Schaaf}, {Scheers}, {Schellart}, {Schoenmakers}, {Schoonderbeek},
  {Serylak}, {Shulevski}, {Sluman}, {Smirnov}, {Sobey}, {Spreeuw}, {Steinmetz},
  {Sterks}, {Stiepel}, {Stuurwold}, {Tagger}, {Tang}, {Tasse}, {Thomas},
  {Thoudam}, {Toribio}, {van der Tol}, {Usov}, {van Veelen}, {van der Veen},
  {ter Veen}, {Verbiest}, {Vermeulen}, {Vermaas}, {Vocks}, {Vogt}, {de Vos},
  {van der Wal}, {van Weeren}, {Weggemans}, {Weltevrede}, {White}, {Wijnholds},
  {Wilhelmsson}, {Wucknitz}, {Yatawatta}, {Zarka}, {Zensus}, \& {van
  Zwieten}}]{van_Haarlem_2013}
{van Haarlem}, M.~P., {Wise}, M.~W., {Gunst}, A.~W., {et~al.} 2013, \aap, 556,
  A2

\bibitem[{{Vedantham} {et~al.}(2012){Vedantham}, {Udaya Shankar}, \&
  {Subrahmanyan}}]{vedantham_et_al2012}
{Vedantham}, H., {Udaya Shankar}, N., \& {Subrahmanyan}, R. 2012, \apj, 745,
  176

\bibitem[{{Voytek}(2015)}]{2015PhDT........65V}
{Voytek}, T.~C. 2015, PhD thesis, Carnegie Mellon University

\end{thebibliography}
    
    \end{document}